\documentclass[twocolumn]{aastex62}

\graphicspath{{./figs/},
}
\usepackage{comment}
\usepackage{amsmath,amssymb}
\usepackage{bm}
\usepackage{soul}

\newcommand{\me}{M_\oplus}
\newcommand{\re}{R_\oplus}
\newcommand{\mjup}{M_\mathrm{Jup}}

\newcommand{\kepler}{\textit{Kepler}}

\newcommand{\gaia}{\textit{Gaia}}

\newcommand{\rsun}{R_\odot}

\newcommand{\au}{\mathrm{AU}}
\newcommand{\yr}{\mathrm{yr}}

\renewcommand{\d}{\mathrm{d}}

\newcommand{\days}{\mathrm{days}}

\newcommand{\cj}{\mathrm{CJ}}
\newcommand{\se}{\mathrm{SE}}

\shorttitle{Are Exo-Jupiter Systems Flat?}
\shortauthors{Masuda et al.}

\begin{document}


\title{
Mutual Orbital Inclinations Between Cold Jupiters and Inner Super-Earths
}

\correspondingauthor{Kento Masuda}
\email{kmasuda@ias.edu}

\author[0000-0003-1298-9699]{Kento Masuda}
\altaffiliation{NASA Sagan Fellow}
\affiliation{Department of Astrophysical Sciences, Princeton University, Princeton, NJ 08544, USA}
\affiliation{Institute for Advanced Study, Princeton, NJ 08540, USA}

\author[0000-0002-4265-047X]{Joshua N.\ Winn}
\affiliation{Department of Astrophysical Sciences, Princeton University, Princeton, NJ 08544, USA}

\author[0000-0003-3309-9134]{Hajime Kawahara}
\affiliation{Department of Earth and Planetary Science, The University of Tokyo, Tokyo 113-0033, Japan}
\affiliation{Research Center for the Early Universe, School of Science, The University of Tokyo, Tokyo 113-0033, Japan}




\begin{abstract}

Previous analyses of Doppler and {\it Kepler} data have found that
Sun-like stars hosting ``cold Jupiters'' (giant planets with $a\gtrsim$~1\,AU)
almost always host ``inner super-Earths'' (1--4\,$R_\oplus$, $a\lesssim$~1\,AU).
Here, we attempt to determine the degree of alignment
between the orbital planes of the cold Jupiters and the inner super-Earths.
The key observational input is the fraction of {\it Kepler} stars
with transiting super-Earths that also have
transiting cold Jupiters.
This fraction depends on both the probability for cold Jupiters to occur
in such systems, and on the mutual orbital inclinations.
Since the probability of occurrence has already been
measured in Doppler surveys, we can use the data to
constrain the dispersion of the mutual inclination distribution. 
We find $\sigma=11.8^{+12.7}_{-5.5}\,\mathrm{deg}$ (68\%
confidence) and $\sigma>3.5\,\mathrm{deg}$ (95\% confidence), where $\sigma$ is the scale parameter of the Rayleigh distribution.
This suggests that
planetary orbits in systems with cold Jupiters tend to be coplanar --- although not quite as
coplanar as those in the Solar System,
which have a mean inclination from the invariable plane of $1.8\,\mathrm{deg}$.
We also find evidence that cold Jupiters have lower mutual inclinations
relative to inner systems with higher transit multiplicity.
This suggests a link between the dynamical excitation in the inner and outer systems.
For example, perturbations from misaligned cold Jupiters may dynamically heat or destabilize systems of inner super-Earths.
 
\end{abstract}




\section{Introduction}\label{sec:intro}

Doppler surveys showed long ago that wide-orbiting giant planets have a broader eccentricity distribution than the planets in the Solar System \citep[e.g.,][]{2006ApJ...646..505B}.
Since planetary systems are thought to form in a ``dynamically cold'' state, with
nearly circular and coplanar orbits, high eccentricities
are usually interpreted as evidence for post-formation perturbations 
due to planet--planet interactions or the influence of a neighboring star \citep[e.g.,][]{1996Sci...274..954R, 1996Natur.384..619W, 1997Natur.386..254H}.
If these systems have been dynamically heated, one might
also expect the mutual inclinations between planetary orbits to have a broader range than the dispersion of a few degrees seen in the Solar System. This prediction
has been difficult to test because 
mutual inclinations are more difficult to measure than eccentricities.

The Doppler method only allows mutual inclinations to be measured directly when planet--planet gravitational interactions
are unusually strong \citep[see, e.g.,][]{2002ApJ...579..455L}. In those few special
cases, all of which involve multiple giant planets, the mutual inclinations have been found to be $\lesssim$\,$10^\circ$.
The Doppler and astrometric techniques have also been used together to demonstrate a $30^\circ$ inclination between two giant planets in the $\nu$\,And system
\citep{2010ApJ...715.1203M}.
Less directly, \citet{2014Sci...346..212D} found evidence for $\approx$40$^\circ$ mutual inclinations in systems featuring a ``warm Jupiter'' (semi-major axis $a=0.1$ to $1\,\mathrm{AU}$) and a ``balmy Jupiter'' ($a>1\,\mathrm{AU}$),
based on the observed tendency for the directions of periapses of the two orbits to
be nearly orthogonal.

The transit method can also provide direct constraints on mutual
inclinations in special cases, as well as indirect constraints based on
observed statistical patterns.
Direct constraints have been obtained from analyses
of the observed variations in transit times and durations in strongly interacting
multi-planet systems \citep[see, e.g.,][]{2012Natur.487..449S,2012Sci...336.1133N},
and a unique example of a planet--planet eclipse \citep{2012ApJ...759L..36H}.
In these cases the mutual inclinations were found to be smaller than about $10^\circ$,
with the exception of Kepler-108, in which two Saturn-mass planets have
orbits inclined by $24^{+11}_{-8}$~degrees \citep{2017AJ....153...45M}.
A less direct method is based on the comparison of transit impact parameters
\citep{2014ApJ...790..146F}. This method has
been used to show that mutual inclinations are typically smaller than a few degrees among the
{\it Kepler} compact multi-planet systems --- except when the innermost
planet has an exceptionally small orbit (with an orbital distance less than
5--6 stellar radii) in which case the inclination dispersion is $\gtrsim$$7^\circ$ \citep{2018ApJ...864L..38D}.

A different indirect technique is based on the observed transit multiplicity function:
the relative numbers of stars with 1, 2, 3, $\cdots$, $N$ detected transiting planets.
All else being equal, flatter planetary systems are more likely to be observed as systems with higher transit multiplicity. However, the transit multiplicity function depends not only on mutual inclinations
but also on the multiplicity function without regard to transits, i.e., the
probability that 1, 2, 3, $\cdots$, $N$ planets exist at all.
For example, the detection of a single transiting planet
might mean only one planet exists, or
that the planet is part of a system of multiple planets on misaligned orbits.
\cite{2018ApJ...860..101Z} broke this degeneracy by analyzing both the transit
multiplicity function and the occurrence of transit timing variations (TTVs), which are less sensitive to mutual inclinations. 
They found a model that successfully explains the observed transit multiplicity function of
{\it Kepler} planetary systems, including the relatively large number of stars with a single detected
transiting planet \citep{2011ApJS..197....8L, 2012ApJ...761...92F, 2012AJ....143...94T, 2012ApJ...758...39J}.
In their model, the mutual inclination dispersion is assumed to decrease with multiplicity:
in systems with 3 or more planets the dispersion is found to be $\lesssim5^\circ$, while for two-planet systems the dispersion is found to be $\sim10^\circ$.

The degeneracy between multiplicity and mutual inclinations can also be broken by combining the results of
transit and Doppler surveys \citep{2012AJ....143...94T, 2012A&A...541A.139F}, which is
the technique employed in this paper.
Specifically, we attempted to determine the distribution of mutual inclinations between two
types of planets which sometimes occur together in the same system:
the typical {\it Kepler} planets with sizes between 1 and $4\,R_\oplus$ and orbital
distances less than $1\,\mathrm{AU}$, which we call ``inner super-Earths'' or simply ``SEs'';
and giant planets on orbits wider than $1\,\mathrm{AU}$, which we call
``cold Jupiters'' or ``CJs''.\footnote{We use the term ``cold'' as a simple 
way to distinguish these planets from the traditionally defined
``hot Jupiters'' ($a<0.1$\,AU) and ``warm Jupiters'' ($0.1\,\mathrm{AU}<a<1$\,AU).
The terminology is not ideal because the actual planet Jupiter, at $5\,\mathrm{AU}$, is colder than most of the known ``cold Jupiters''. This may be why \citet{2014Sci...346..212D} preferred the name
``balmy Jupiters''.}

We chose to focus on SEs and CJs because the necessary information has become available only recently,
and because previous studies have found an unexpectedly strong link between these two
populations \citep{2018AJ....156...92Z, 2019AJ....157...52B}.
Both transit and Doppler studies have shown that SEs exist around one-third of Sun-like
stars, and that SEs are accompanied by CJs about one-third of the time --- implying that
one in nine Sun-like stars has both a SE and a CJ.
This is interesting because $1/9$ is similar to
the overall occurrence rate of CJs found in Doppler surveys, unconditioned on the presence of SEs
\citep{2008PASP..120..531C, 2011arXiv1109.2497M, 2019ApJ...874...81F}.
The implication is that nearly all CJs are accompanied by SEs.
Such a strong association must be a clue about the
formation and evolution of these systems.
It seems to contradict the scenario modeled by
\citet{2015ApJ...800L..22I} in which CJs block the migration of SEs,
as well as the scenario of \citet{2017A&A...604A...1O} in which CJs
inhibit the formation of SEs by blocking the flux of pebbles from the outer regions of the protoplanetary
disk. In either scenario, it would be rare to find both an inner SE and a CJ.
On the other hand, a positive correlation would be expected if the close-in SEs form {\it in situ} from a massive disk
\citep{2013MNRAS.431.3444C, 2014ApJ...790...91S}.

Given the strong correlation between SEs and CJs, it is also of interest to understand their mutual
inclinations.
This serves as a probe of the degree of dynamical excitation, complementary to the orbital eccentricity.
The mutual inclination is also important
for understanding how the gravitational perturbations from the CJ have affected the properties
of the inner SEs.
Several theoretical works have shown 
that CJs on inclined orbits can dynamically heat the inner SE systems,
increasing the inclination dispersion of the inner system
or even destabilizing the orbits and reducing the multiplicity through collisions and ejections
\citep{2017AJ....153...42L, 2017AJ....153..210H, 2017MNRAS.464.1709G, 2017MNRAS.467.1531H, 2017MNRAS.468.3000M, 2017MNRAS.468..549B, 2018MNRAS.478..197P}.
Indeed, some evidence has already been found that a subset of these compact systems
are dynamically hot \citep{2016PNAS..11311431X, 2019AJ....157...61V,2018ApJ...860..101Z, 2018ApJ...864L..38D, 2019AJ....157..198M}. 
Knowledge of how, and how often, the inner systems are perturbed is
essential for understanding the diversity of close-in SE systems.

This paper is organized as follows. Section~\ref{sec:simple} presents the simple and approximate calculation that
inspired the more rigorous and thorough examination that is presented in the rest of the paper.
Section \ref{sec:framework} lays out the mathematical framework of the more complex
calculation.  Section \ref{sec:occ_doppler}
describes one of the key observational
inputs: the occurrence rate of CJs around stars with inner SEs.
Section \ref{sec:sesample} describes another crucial input:
the construction of a sample of stars with transiting inner SEs,
and a subsample that also has transiting CJs.
Section \ref{sec:results} presents the results for the mutual inclination distribution and its
dependence on the transit multiplicity of the inner system.
Section \ref{sec:tests} relates some tests we performed on the robustness of the results to reasonable
variations in the analysis procedure.
After our analysis was completed, we learned of an analysis
with a similar goal by \citet{2019AJ....157..248H}.
Section \ref{sec:compare} presents a brief comparison between our work
and theirs.
Finally, Section \ref{sec:discussion} summarizes the results and relates them to other observations and theories for the
architecture of compact systems of SEs.

\section{Simple Calculation}
\label{sec:simple}

The \kepler\ database includes about 1{,}000 stars with at least one confirmed transiting inner super-Earth.
As we will see in Section \ref{sec:occ_doppler}, the available
Doppler data imply that about one-third of these SE systems should
also harbor a cold Jupiter.
If the SE and CJ orbital orientations were uncorrelated,
then the transit probability for a CJ would be $R_\star/a$, where $R_\star$ is the stellar
radius and $a$ is the orbital distance.  If the period $P$ is longer than
the 4-year duration of the prime {\it Kepler} mission, the probability to transit during the mission
is reduced by a factor of about $4\,\mathrm{yr}/P$.
This transit probability is about $1/500$ when averaged over the range of orbital periods of the observed CJs.
Thus, if the orbits were uncorrelated, the expected number of stars with transiting SEs 
and transiting CJs would be about $1000/3/500$ or $2/3$.
However, there are 3 known cases of transiting CJs in systems with inner SEs, as detailed
in Section \ref{sec:sesample}.  This is 4.5 times larger than the expected value, indicating
that the orbital orientations of SEs and CJs are correlated.

We can use this observation to place a rough constraint on the typical mutual inclination $\theta$ between the
inner SEs and the outer CJ.  Whenever an outer planet (with orbital distance $a_{\rm out}$) and an inner planet ($a_{\rm in}$) have a mutual inclination
larger than $R_\star/a_{\rm in}$, and the inner planet is transiting, then the 
transit probability of the outer planet is larger than $R_\star/a_{\rm out}$ by a factor of approximately
$1/\sin\theta$ \citep{2010arXiv1006.3727R}.
By setting this geometrical factor equal to the observed $4.5\times$ enhancement in the occurrence of transiting CJs in
SE systems, we find
\begin{equation}
\frac{1}{\sin\theta} \approx 4.5 \longrightarrow \theta \approx 13^\circ.
\end{equation}

Moreover, in all 3 cases in which transiting CJs have been found in this sample, there is more
than one transiting inner SE. This is despite the fact that stars with multiple transiting SEs
constitute only $1/3$ of the sample. Thus, considering only the sample of multi-transiting SE systems, the observed occurrence of
transiting CJs is larger than would be expected for uncorrelated orbits by a factor
of $3\times 4.5$ or 13.5.  Using the equation above with 13.5 instead of 4.5,
the typical mutual inclination is approximately $4^\circ$.
As for the other $2/3$ of the sample, consisting of single-transiting SE systems, the lack of any
detections of transiting CJs is compatible with the
assumption of uncorrelated orbits.

In short, these rough calculations indicate that the orbits of CJs have a typical inclination
of $\sim$$10^\circ$ with respect to inner SEs, and also that the inclination is smaller when
the inner system is composed of multiple transiting SEs.
The rest of this paper describes our effort to perform these calculations more rigorously
and understand the limitations and implications of the results.

\section{Complex Calculation}\label{sec:framework}

We denote by $\mathcal{S}$ the sample of 
\kepler\ stars with transiting inner SEs.
The expected number of transiting CJs in this sample depends
on both the probability that CJs exist in a system with SEs, as well
as the probability distribution $p(\theta)$ for the
angle between the orbits of the CJ and the SEs.
We assume that $p(\theta)$ can be well modeled as a
von Mises--Fisher distribution \citep{fisher}
\begin{equation}
	p(\theta|\kappa)\,\d\theta \propto \exp(\kappa\cos\theta)\,\sin\theta\,\d\theta, \quad 0 \leq \theta \leq \pi.
\end{equation}
Here, $\kappa>0$ is the concentration parameter, which is small for nearly isotropic distributions and
large for distributions that are sharply peaked around $\theta=0$.
We prefer to parameterize the distribution in terms of $\sigma \equiv \kappa^{-1/2}$,
because as $\sigma\rightarrow 0$ the distribution approaches the familiar Rayleigh distribution with scale parameter $\sigma$, for which the mode is $\sigma$, the mean is $(\pi/2)^{1/2}\sigma\approx1.25\sigma$, and the root-mean-square value is $\sqrt{2}\sigma$.
We will compute $p(n_{\rm tCJ}|\sigma)$, the probability distribution for the number of transiting CJs in the sample, and compare it to $n_{\rm tCJ,\,obs}(\mathcal{S})$,
the observed number of transiting CJs in the sample.
In this way, we will determine the relative likelihood of different values of $\sigma$.

We calculate $p(n_{\rm tCJ}|\sigma)$ via Monte Carlo simulation:
\begin{enumerate} 

\item We assign one CJ to a fraction $Z$ of randomly chosen stars in $\mathcal{S}$. This subset
is denoted $\mathcal{S}_{\rm CJ}$. The chosen fraction $Z$ is based on
the previous work of \citet{2018AJ....156...92Z} and \citet{2019AJ....157...52B},
as described in Section~\ref{sec:occ_doppler}.
Each CJ is randomly assigned an orbital period $P_{\cj,i}$ and eccentricity $e_{\cj,i}$ based on the results of Doppler surveys, in a manner to be discussed
in Section \ref{sec:occ_doppler}. The argument of periastron $\omega_{\cj,i}$ is randomly drawn from a uniform distribution.

\item Each CJ in $\mathcal{S}_{\rm CJ}$ is also assigned an orbital inclination $I_{\cj,i}$ relative
to the line of sight. This is computed by drawing a mutual inclination $\theta_i$ from the probability
distribution $p(\theta|\sigma)$, and combining it with the mean orbital inclination $I_{\se,i}$ of the inner
system of SEs:
\begin{equation}
	\label{eq:cosicj}
	\cos I_{\cj,i}=\cos\theta_i\cos I_{\se,i} + \sin\theta_i\sin I_{\se,i}\cos\phi,
\end{equation}
where $\phi$ is a phase angle randomly sampled from a uniform distribution.
The relevant geometry is shown in Figure \ref{fig:geo}.
The figure shows that the last term in the preceding equation should be
$\cos (\phi_{\rm LOS}-\phi_{\cj})$, which has the same distribution as the cosine of
a uniformly distributed random variable under the assumptions that
$\phi_{\rm LOS}$ and $\phi_{\cj}$ are themselves uncorrelated and uniformly distributed.

\begin{figure*}[ht]
	\epsscale{0.9}
	\plotone{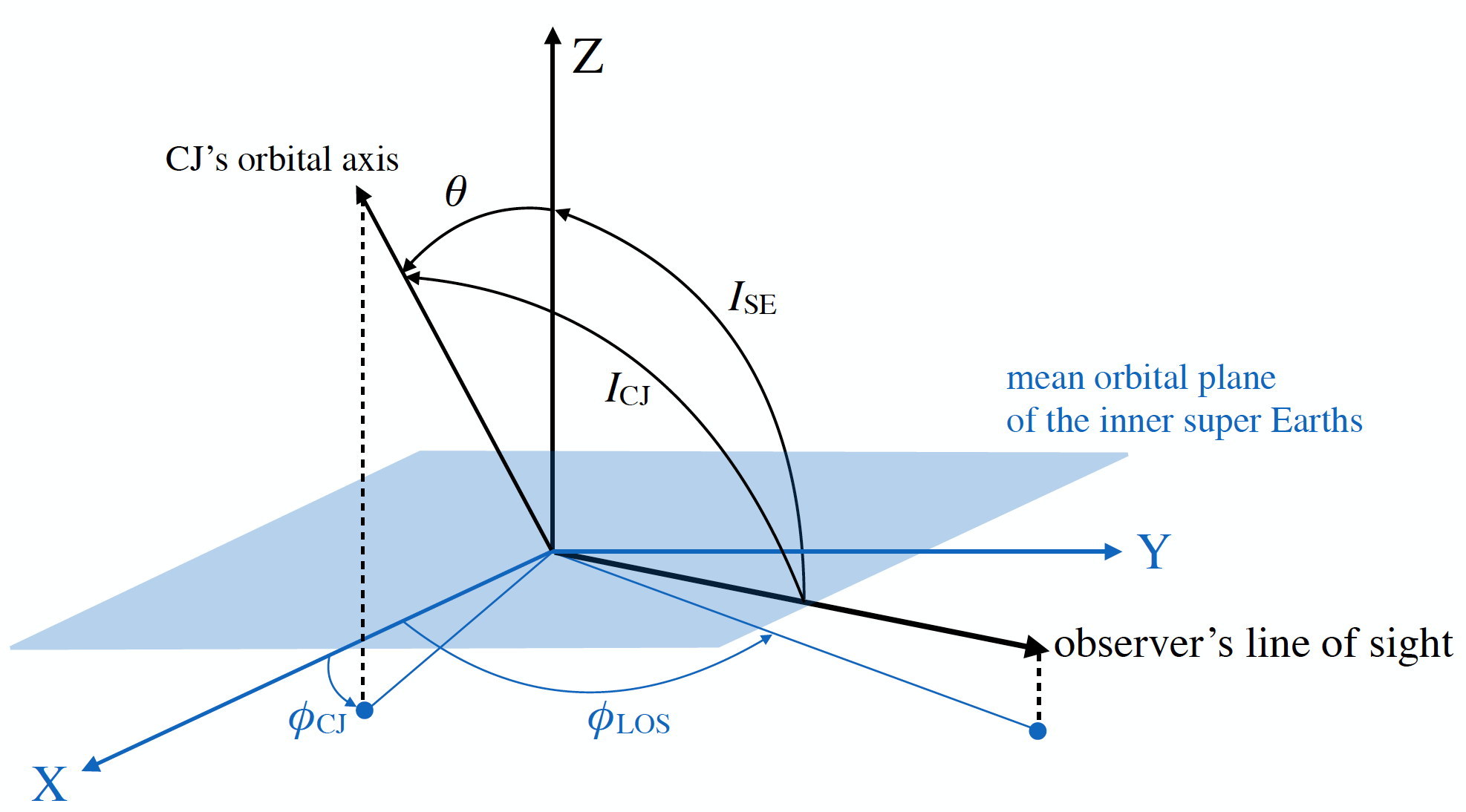}
	\caption{Geometry of the system. The angle $\theta$ is the mutual inclination between the orbit
	of the cold Jupiter and the mean orbital plane of the inner super-Earths.
	We are ignorant about the phase angles $\phi_{\cj}$ and $\phi_{\rm LOS}$.}
	\label{fig:geo}
\end{figure*}

We estimate $I_{\rm \se}$ using the reported values of
orbital inclination that were determined by fitting the transit light curves.
When there is more than one transiting inner planet, we set $I_{\se}$ equal to the
mean orbital inclination of all the planets.\footnote{Strictly speaking, there is ambiguity in this procedure
because of the degeneracy between $I$ and $180^\circ - I$ when fitting transit light curves.
We always choose values of $I$ to be less than 90$^\circ$,
motivated by the finding of \citet{2014ApJ...790..146F}
that wider-orbiting planets tend to have higher transit impact parameters,
indicating that multiple transiting SEs tend to have mutual inclinations
of at most a few degrees.
In any case, this choice only affects $p(n_{\rm tCJ}|\sigma)$ for $\sigma \lesssim$ a few degrees,
and our conclusions will not be sensitive to the difference in this region.}

\item We check on whether or not each of the assigned CJs exhibits transits.
We compute the impact parameter
\begin{align}
	\notag
	b_{\cj,i}
	&={a_{\cj,i}(P_{\cj,i}, M_{\star,i})\,\cos I_{\cj,i} \over R_{\star,i}}\,{{1-e_{\cj,i}^2}\over{1+e_{\cj,i}\sin\omega_{\cj,i}}},
\end{align}
and set $n_{\mathrm{tCJ},i}=0$ if $b_{\cj,i}>1$. If $b_{\cj,i}<1$, we also
compute the total duration $T_{\mathrm{obs},i}$ of observations of the host star,
by multiplying the timespan of the {\it Kepler} data by the duty cycle.
When $T_{\mathrm{obs},i}>P_{\cj,i}$, we set $n_{\mathrm{tCJ},i}=1$.
When $T_{\mathrm{obs},i}<P_{\cj,i}$, the transit of the CJ might fall outside the observing duration; we set $n_{\mathrm{tCJ},i}=1$ with probability $T_{\mathrm{obs},i}/P_{\cj,i}$ and $n_{\mathrm{tCJ},i}=0$ otherwise.
Finally, we compute
\begin{equation}
	n_{\rm tCJ|\sigma} = \sum_{i \in \mathcal{S}_{\rm CJ}} n_{\mathrm{tCJ},i}.
\end{equation}
Here we have assumed that the transits of the simulated CJs are always detected whenever they occur.
As explained in Section \ref{sec:sesample}, we will choose the sample $\mathcal{S}$ so that this assumption is justified.
\end{enumerate} 
For each value of $\sigma$, we repeat these steps for different values of $Z$ that are compatible
with the observational uncertainty. The Monte Carlo simulation is repeated
for a range of possible values of $\sigma$, thereby delivering
the desired probability distribution $p(n_{\rm tCJ}|\sigma)$.
Finally, we compute the likelihood function $\mathcal{L}(\sigma)\equiv p(n_{\rm tCJ,obs}|\sigma)$
in order to draw inferences about the mutual inclination distribution.

\section{Occurrence Rate of Cold Jupiters in Super-Earth Systems}\label{sec:occ_doppler}

\begin{deluxetable*}{lcccc}[!ht]
\tablecaption{Estimated occurrence rate of cold Jupiters (CJs) around stars with inner super-Earths (SEs), drawn from the literature.\label{tab:cjocc}}
\tablehead{
\colhead{} & \colhead{Definition of SEs} & \colhead{Definition of CJs} & \colhead{Fraction $Z$}
}
\startdata
\multicolumn{4}{l}{{\it \citet{2018AJ....156...92Z}, FGK stars}}\\
\quad Doppler & $m\sin i<20\,\me$, $P<400\,\days$ & $m\sin i>95\,\me$ ($0.3\,\mjup$), $P\gtrsim1\,\mathrm{yr}$ 
& $10/32=(32\pm8)\%$\\
\quad transit & $r=1$--$4\,\re$, $P<400\,\days$ & $m\sin i>95\,\me$ ($0.3\,\mjup$), $P\gtrsim1\,\mathrm{yr}$ & $7/22=(33\pm10)\%$\\
\quad combined & $r=1$--$4\,\re$ or $m\sin i<20\,\me$, $P<400\,\days$ & $m\sin i>95\,\me$ ($0.3\,\mjup$), $P\gtrsim1\,\mathrm{yr}$ & $17/54=(32\pm6)\%$\\
\tablebreak
\multicolumn{4}{l}{{\it \citet{2019AJ....157...52B}, FGKM stars}}\\
\quad Doppler & $m\sin i=1$--$10\,\me$, $a<0.5\,\au$
& $m=0.5$--$20\,\mjup$, $a=1$--$20\,\au$ ($P\simeq1$--$90\,\mathrm{yr}$) & $34^{+11}_{-10}\%$\\
\quad transit & $r=1$--$4\,\re$, $a<0.5\,\au$
& $m=0.5$--$20\,\mjup$, $a=1$--$20\,\au$ ($P\simeq1$--$90\,\mathrm{yr}$) & $41^{+10}_{-10}\%$\\
\quad combined & $r=1$--$4\,\re$ or $m\sin i=1$--$10\,\me$, $a<0.5\,\au$
& $m=0.5$--$20\,\mjup$, $a=1$--$20\,\au$ ($P\simeq1$--$90\,\mathrm{yr}$) & $(39\pm7)\%$\\
\quad combined 2 & $r=1$--$4\,\re$ or $m\sin i=1$--$10\,\me$, $a<0.5\,\au$
& $m=0.5$--$20\,\mjup$, $a=1$--$10\,\au$ ($P\simeq1$--$30\,\mathrm{yr}$) & $(38\pm7)\%$\\
\quad combined 3 & $r=1$--$4\,\re$ or $m\sin i=1$--$10\,\me$, $a<0.5\,\au$
& $m=1$--$13\,\mjup$, $a=1$--$10\,\au$ ($P\simeq1$--$30\,\mathrm{yr}$) & $(34\pm7)\%$\\
\enddata
\end{deluxetable*}

\citet{2018AJ....156...92Z} and \citet{2019AJ....157...52B} independently analyzed the occurrence rate of Doppler CJs around stars with inner SEs, based on samples of SE systems
drawn from both the {\it Kepler} transit survey and from precise Doppler surveys.
Their key results are summarized in Table \ref{tab:cjocc}. There are a number of differences between the two analyses, including the types of host stars considered, the adopted
definitions of SEs and CJs, and the treatment of candidate CJs for which the available
Doppler data do not span a full orbit. Nevertheless, the occurrence rates derived by these authors are consistent with each other. The rates derived from the sample of Doppler SEs and
from the sample of transiting SEs do not show significant differences, either.
For our purpose, we adopted $Z=(34\pm7)\%$, based on the results of
\citet{2019AJ....157...52B} for the narrowest ranges of mass and semi-major axis they
considered (i.e., the smallest amount of extrapolation).

In constructing our sample $\mathcal{S}$, we needed to match the definitions
of CJs and SEs as closely as possible to the definitions that were adopted
by \citet{2019AJ....157...52B}.
For this reason, we defined inner SEs as planets with radii $r=1$--$4\,\re$ and orbital periods $P<130\,\days$ ($a<0.5\,\au$ for a solar-mass star).\footnote{Although \citet{2018AJ....156...92Z} included SEs with periods as long as $400\,\days$, the difference is minor because there are very few \kepler\ transiting SEs with periods
between 130 and 400~days.}
A complication that arose when defining CJs is that \citet{2019AJ....157...52B} used a mass-based
definition while we needed to use a radius-based definition.
Fortunately, the mass--radius relation for giant planets is well understood:
the radius hardly changes at all with mass, for objects between $0.3$ and $30\,\mjup$.
The smallest known planet in this mass range with both well-determined mass and radius has a radius of $7.7\pm0.4\,\re$ \citep[K2-60 b,][]{2017AJ....153..130E}. 
Therefore, we defined CJs as planets with $r>7.5\,\re$ and $P=1$ to $30\,\yr$.
An unavoidable problem with this definition is that it includes
brown dwarfs with masses $>13\,\mjup$, which were not always counted as planets
in the Doppler surveys, but this is a minor problem because brown dwarfs are much
less common than giant planets in this period range \citep[e.g.,][]{2006ApJ...640.1051G}.
Our definition also inadvertently includes
any low-mass, low-density planets that are
less massive than about $0.3\,\mjup$ and larger than $7.5\,\re$.
Such objects are known to exist
in $<1\,\mathrm{yr}$ orbits, but they are less common than ordinary SEs by at least
an order of magnitude \citep{2018AJ....155...89P}. For our analysis we had
to assume that their occurrence rate is also much lower for orbital periods longer than $1\,\mathrm{yr}$, than that of traditional giant planets (i.e., 1/3 around SEs). If this assumption is mistaken, then our analysis will underestimate the typical mutual inclination.

Next, we needed to assign orbital periods and eccentricities to the simulated CJs.
Since previous studies have concluded that
almost all CJs are associated
with inner SEs, the properties of CJs around SEs should be similar to the properties
of CJs in general that have been derived from Doppler surveys.
\citet{2018AJ....156...92Z} verified that this is the case.
To model the period distribution, we relied on the results of
of \citet{2019ApJ...874...81F}, based on long-term surveys with the High
Accuracy Radial-velocity Planet Searcher (HARPS, \citealt{2011arXiv1109.2497M}). The simplest
model for the 
period distribution that they found to be consistent with the population of giant planets with $P\gtrsim100\,\days$ is
\begin{equation}
	\label{eq:f_f19}
	p(\ln P)\,\mathrm{d}\ln P \propto \mathcal{N}(\ln P_0, \sigma_{\ln P})\,\mathrm{d}\ln P,
\end{equation}
where $\mathcal{N}(\mu, \sigma)$ denotes a normal distribution with mean $\mu$ and
dispersion $\sigma$, and the two constants are $P_0=(919\pm105)\,\mathrm{days}$ and $\sigma_{\ln P}=1.46$ (see their Figure 3).
For the probability distribution of orbital eccentricity, we followed
\citet{2013MNRAS.434L..51K} by adopting a beta distribution,
\begin{equation}
p(e)\,\mathrm{d}e \propto e^{a-1} (1-e)^{b-1},
\end{equation}
with $a=1.12$ and $b=3.09$.

Our model does not allow a star to have 
more than one wide-orbiting giant
planet. The possible multiplicity of CJs is often ignored
in studies of planet occurrence from Doppler surveys, and was handled
differently
by \citet{2018AJ....156...92Z} and \citet{2019AJ....157...52B}.
The former group counted the number of stars with at least one CJ, while the latter group
included only the innermost CJs and verified that their results
were not sensitive to which of the inner or outer planets were included.
By ignoring the possibility of a second CJ, our analysis underestimates
the probability that a star will have a transiting CJ, but we expect
this to be a minor error because
any more distant CJs will have lower transit probabilities --- and as
we will see later, no star in our sample has more than one transiting CJ.
This possible source of systematic error will be addressed
further in Section \ref{ssec:multicj}.

\section{The Sample of Stars with Transiting Super Earths}\label{sec:sesample}

Here we define the sample $\mathcal{S}$ of stars hosting close-in SEs. As discussed in the previous section, the sample needs to satisfy the following criteria:
\begin{enumerate}
\item The stars have reliably detected SEs similar to the ones studied by \citet{2018AJ....156...92Z} and  \citet{2019AJ....157...52B}. 
\item The detection of a transiting CJ would be straightforward,
consistent with the assumption that all transiting CJs around the stars would
have been detected.
\item The stars are similar in their basic properties
to the ones studied by \citet{2018AJ....156...92Z} and
\citet{2019AJ....157...52B}.
\end{enumerate}

We began with the list of 177{,}911 {\it Kepler}
stars presented by \citet{2018ApJ...866...99B}, for which stellar radii were well constrained based on the parallax from the \gaia\ Data Release 2 \citep{2018A&A...616A...1G}
and the effective temperature determined from various sources.
For some of the stars, we adopted the improved (spectroscopic) determinations of effective
temperature from the work of \citet{2019AJ....157..218K}, and updated
the determination of the stellar radius.
These stars were cross-matched with the {\it Kepler} Object of Interest (KOI) catalog, Data Release 25 \citep{2018ApJS..235...38T}, and the planetary radii $r$ were assigned based on the radius ratio in the catalog and the updated stellar radii.

We defined the sample $\mathcal{S}$ of \kepler\ stars with inner SEs based on the following criteria:
\begin{enumerate}
\item The star is not associated with any ``false positive'' planets reported in the
KOI catalog.
\item The star has at least one confirmed inner SE, i.e., one
planet with $P<130\,\days$ and $1<r/\re<4$. We also imposed a restriction
on the transit impact parameter, $b<0.9$, to eliminate cases in which
the planet radius cannot be measured reliably.
\item The star has no detected transiting planets with $P<130\,\days$ and $r>4\,\re$. This is because of the evidence that stars with hot or warm giant planets may have higher occurrence rates of CJs \citep{2016ApJ...821...89B}. This criterion removes only a small number
of stars because hot/warm giant planets are intrinsically rare.
\item The transit depth expected for a $6\,\re$ planet is at least 10 times larger than the Combined Differential Photometric Precision \citep[CDPP,][]{2010ApJ...713L..79K} computed for the $15\,\mathrm{hr}$ timescale. 
This is to ensure that transiting CJs would have been easily detected, 
although this condition was almost always satisfied because the star was already
required to have a detected transiting SE.
Only 21 stars were dropped at this stage.
\end{enumerate}
The resulting sample consists of 1{,}046 stars hosting a total of 1{,}421 SEs.
Among them, 651 stars have a single transiting SE,
and 395 stars have multiple transiting planets including SEs as defined above.
Our main analysis is based only on this sample of stars with confirmed transiting SEs,
although in Section \ref{ssec:candidates} we discuss how the results would
change if we include stars hosting candidate transiting SEs.

\begin{figure*}[ht]
\epsscale{0.92}
\plotone{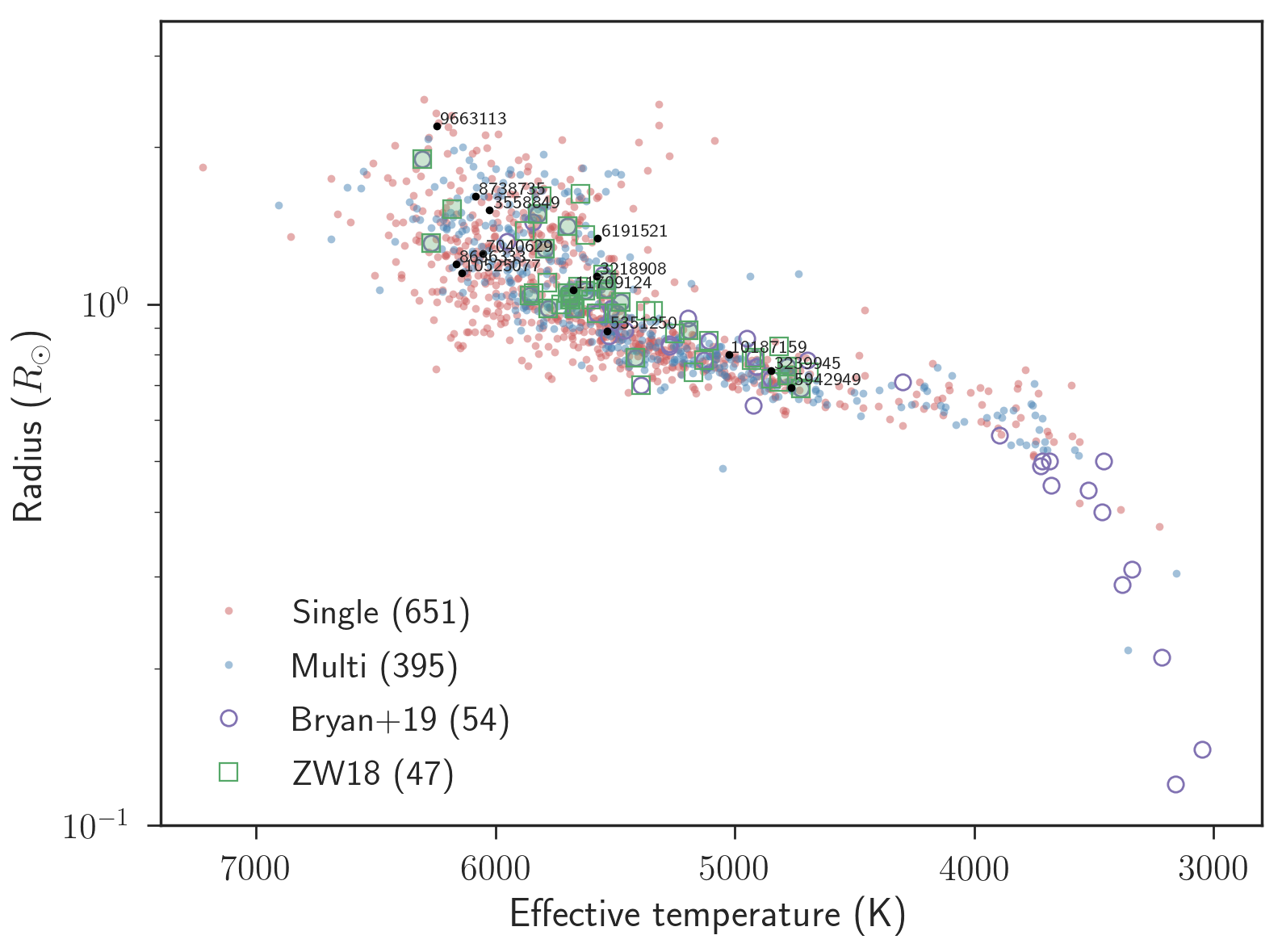}
\caption{The radius and effective temperature of the stars hosting SEs from our sample and two previously
studied samples.
Red and blue dots represent stars from our sample with a single transiting planet
and with multiple transiting planets, respectively. Purple open circles and green open squares are
super-Earth hosts analyzed by \citet{2019AJ....157...52B} and \citet{2018AJ....156...92Z}, respectively. Black dots represent all the \kepler\ stars for which transits of both a long-period giant planet
and an inner super-Earth have been detected
(Table \ref{tab:cjdata}).}
\label{fig:comparison_hr}
\end{figure*}

\begin{figure*}[ht]
\epsscale{0.55}
\plotone{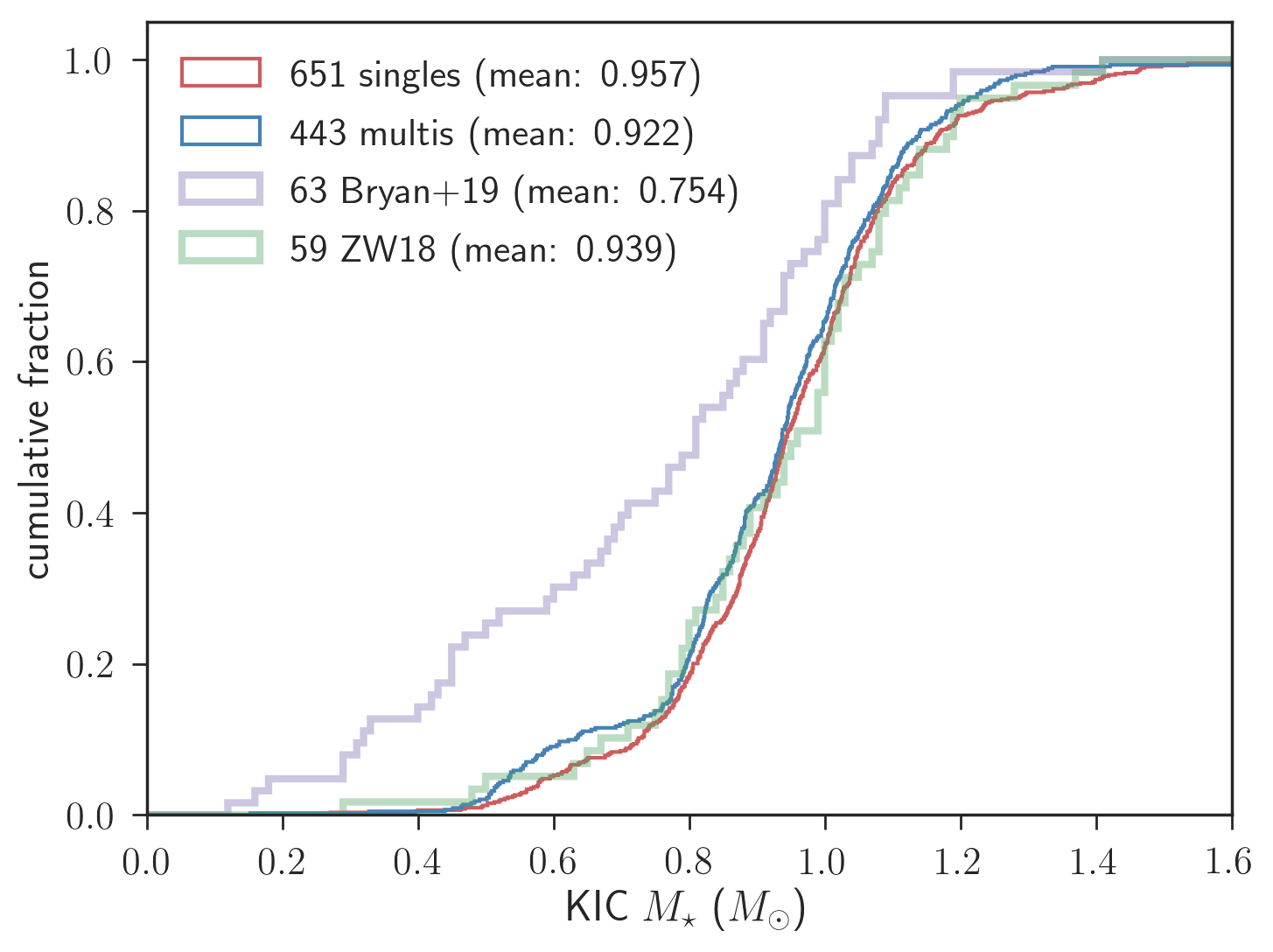}
\plotone{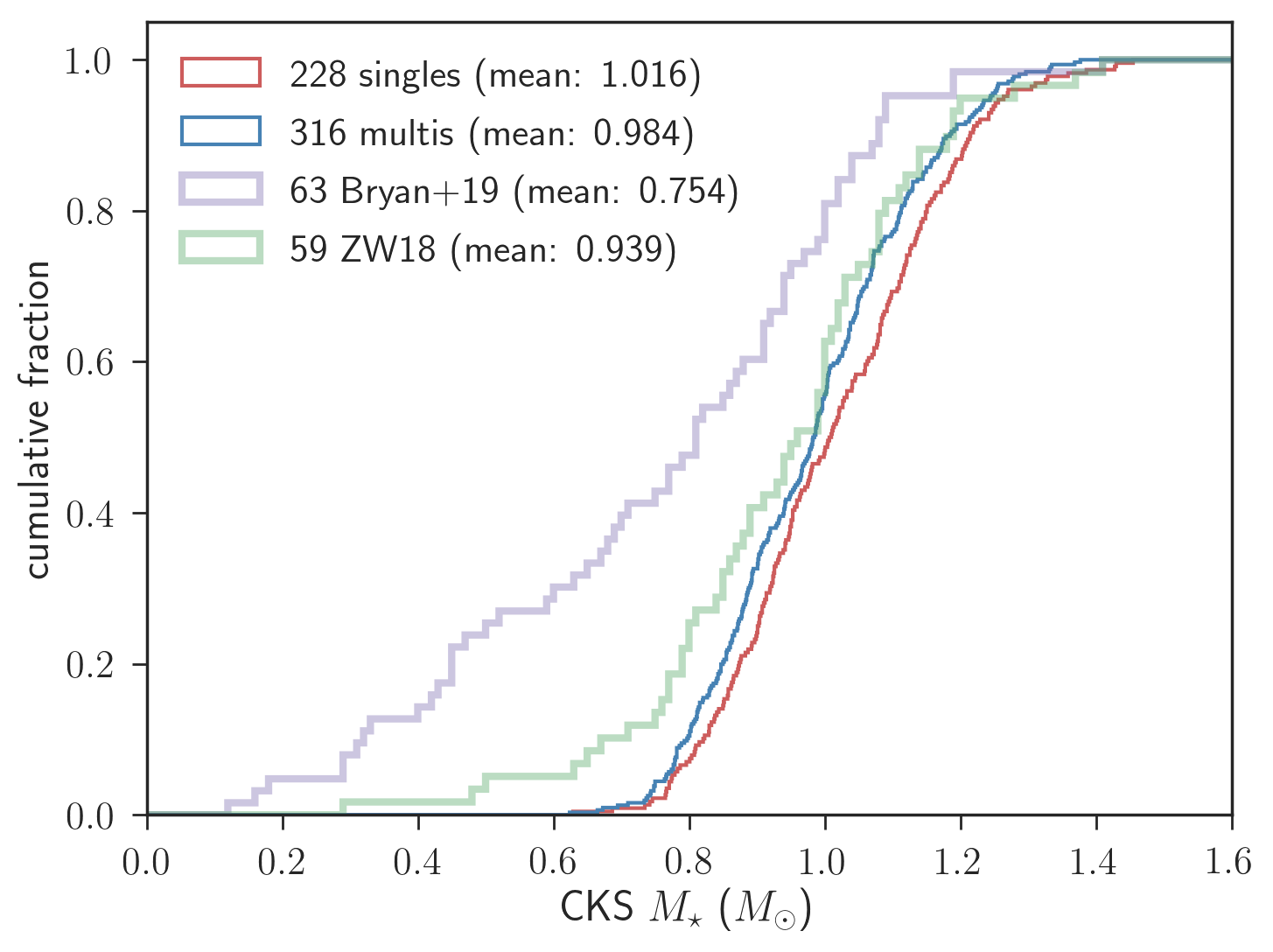}
\plotone{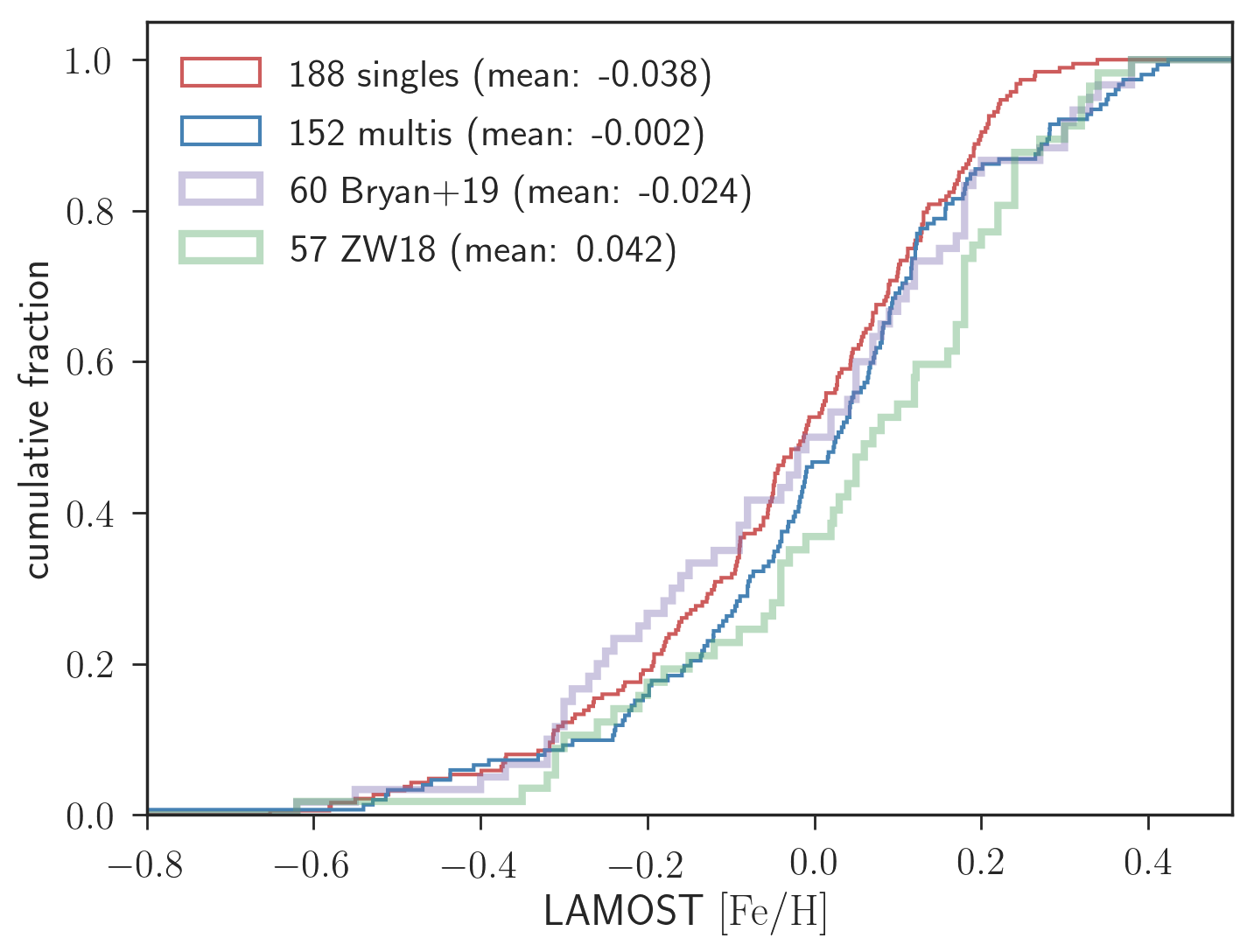}
\plotone{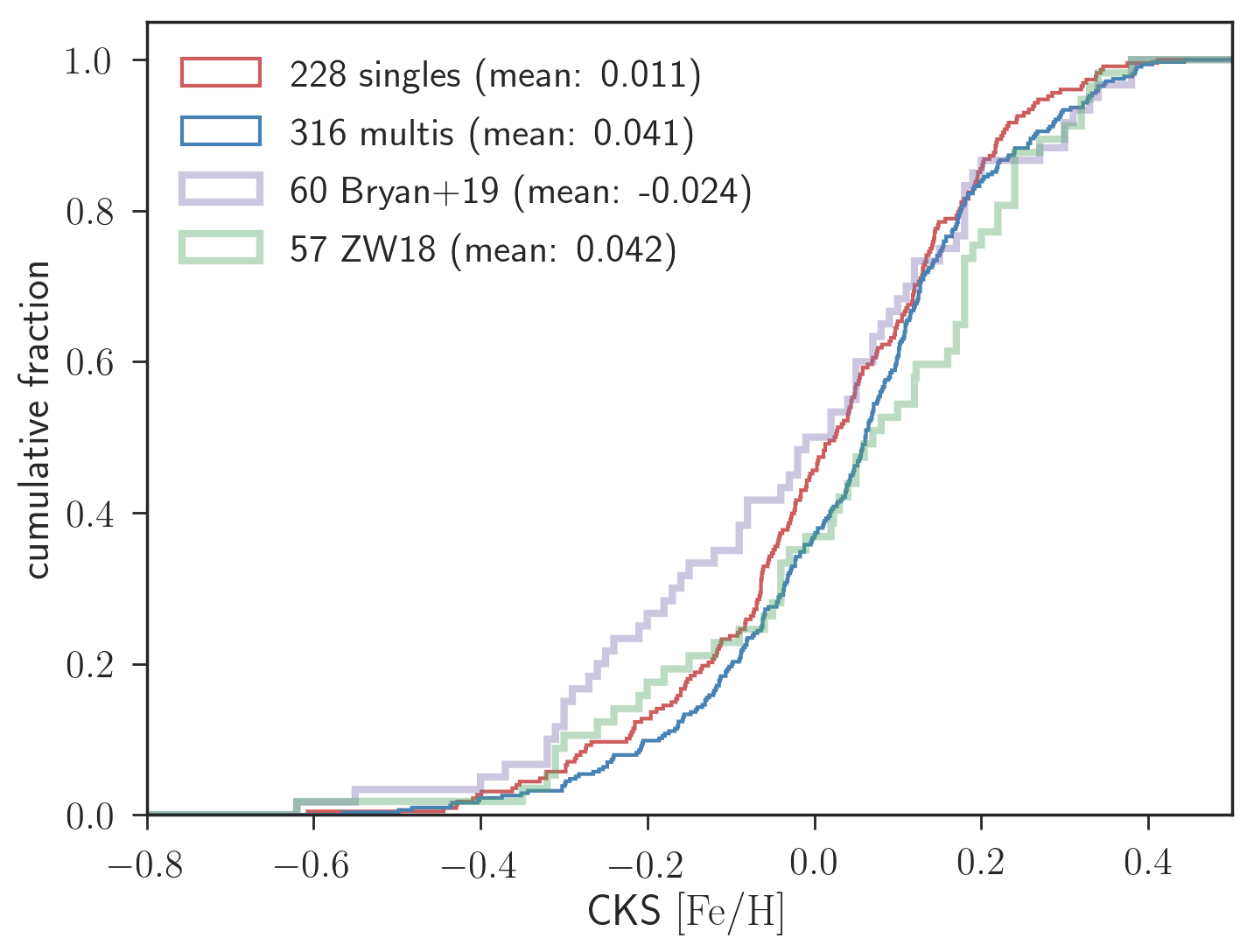}
\caption{Distributions of stellar mass and [Fe/H] for the stars in our
sample and the previously analyzed samples
of \citet{2018AJ....156...92Z} and \citet{2019AJ....157...52B}.
The data for the previously analyzed samples were taken
from the NASA exoplanet archive.
Those for our sample stars are from difference sources: {\it Top left} --- stellar masses from
the {\it Kepler} Input Catalog.
{\it Top right} --- stellar masses from the California \kepler\ survey \citep[CKS;][]{2017AJ....154..107P, 2017AJ....154..108J}. {\it Bottom left} --- [Fe/H] from LAMOST \citep{2012RAA....12.1197C, 2015RAA....15.1095L} DR 4. {\it Bottom-right} --- [Fe/H] from the CKS.}
\label{fig:comparison_hist}
\end{figure*}

Figures \ref{fig:comparison_hr} and \ref{fig:comparison_hist} compare the stellar properties of our sample with those of the stars analyzed
by \citet{2018AJ....156...92Z} and \citet{2019AJ....157...52B} (collected
from the NASA exoplanet archive\footnote{\url{https://exoplanetarchive.ipac.caltech.edu}.  A few stars for which no information was available were ignored.}).
The figures show that \citet{2018AJ....156...92Z} focused on FGK stars, while $\approx 20\%$ of the sample of \citet{2019AJ....157...52B} were M dwarfs.
In addition, the stars from the sample of \citet{2019AJ....157...52B} have a lower mean iron abundance than those in the sample of \citet{2018AJ....156...92Z}, by $0.06\,\mathrm{dex}$. Nevertheless, the two groups derived occurrence rates for CJs outside close-in SEs that are consistent with each other,
as summarized in Table \ref{tab:cjocc}.
The properties of our sample overlap with those of the two previous samples.
We also see no significant difference between the properties of stars with single transiting planets
and those with multiple transiting planets.

\subsection{How Many of Them Have Transiting CJs?}\label{ssec:sesample_tcj}

We consulted the catalog of \citet{2019AJ....157..218K} to identify the stars in our sample $\mathcal{S}$
that have transiting CJs that meet the criteria
$r>7.5\,\re$ and $1\,\mathrm{yr}<P<30\,\mathrm{yr}$.
The catalog includes all of the long-period transiting planet candidates known from the prime \kepler\ mission. We found $n_{\rm tCJ,obs}=3$ for the confirmed SE sample: 
the CJs around KIC 3218908 and 11709124 were originally reported by \citet{2016ApJ...822....2U} and the CJ around KIC 3239945 was validated by \citet{2016ApJ...820..112K}.
These are listed in Table \ref{tab:cjdata} in boldface type.
Although two of them show only one transit in the \kepler\ data, their transit durations combined with stellar radii indicate that their periods are in the range of interest.
They all have three or more inner transiting planets, and they were all detected
both by visual inspection \citep{2016ApJ...822....2U} and through an
automated search \citep{2016AJ....152..206F}. 

We note that KIC 3218908 was classified as a photometric binary by \citet{2018ApJ...866...99B} because the calculated stellar radius of $1.22\,\rsun$ was larger than expected for the estimated effective temperature of $4938\,\mathrm{K}$. However, a more recent spectroscopic determination of the effective temperature is $5578\,\mathrm{K}$ \citep{2019AJ....157..218K}, placing the star closer to the main sequence (see Figure \ref{fig:comparison_hr}) and making this star less anomalous. In any case, the system would still meet our criteria for the ``CJ+SE" sample even if the transit signal was diluted by an unresolved companion with a similar luminosity.

The super-Earth sample $\mathcal{S}$ contains about 1{,}000
stars and its composition would not change substantially if we made
small changes to the selection criteria.
However, since the number of transiting CJs in this sample is only 3, we need to be more careful about the effects on the results of any minor changes
in the selection criteria.
To check on this, Table \ref{tab:cjdata} has a complete listing
of all the stars known to have both long-period transiting planets and inner transiting planets.
Some of the entries are not counted in $n_{\rm tCJ,obs}$ 
because the outer planets are too small.
Those dropped all have $r\lesssim6.4\,\re$ (KIC 10187159); empirically, such planets are
very likely to be less massive than $0.3\,\mjup$, and thereby do not
fall inside the definitions of CJs for which the conditional occurrence is known.
The italicized entries are not counted in $n_{\rm tCJ,obs}$ 
because the star is not included in the SE sample $\mathcal{S}$.
The transiting CJs around KIC 9663113 and KIC 6191521 were not included in $\mathcal{S}$
because the inner transiting planets are giant planets.
Thus, $n_{\rm tCJ,obs}$ in the confirmed SE sample appears to be robust
against small changes in the definitions of close-in SEs and CJs.

We also checked for any stellar companions found by the Robo-AO \kepler\ survey \citep{2018AJ....156...83Z}.
Companions within 4~arcsec were detected for KIC 9663113 (KOI-179) and KIC 8636333 (KOI-3349), but the associated corrections to the planet radii are either negligibly small (if the planet
orbits the primary star), or so large that the system becomes physically implausible or irrelevant to our purpose (if the planet orbits the secondary star).
The other stars in Table \ref{tab:cjdata} have no detected companions.

We emphasize that this list needs to be complete only for Jupiter-sized transiting planets in our sample of KOI stars with close-in SEs, on which our analysis is solely conditioned. This is a much less stringent requirement than performing a complete search of long-period transiting planets around generic \kepler\ stars. Our search need not be complete for planets smaller than about $7\,\re$, nor for planets around stars without transiting SEs (i.e., the vast majority of the \kepler\ stars).
The light curves of KOI stars have been thoroughly examined via both visual inspection \citep{2016ApJ...822....2U} and automated algorithms \citep{2016AJ....152..206F, 2019AJ....157..248H},
and the resulting samples of transiting Jupiter-sized planets around confirmed KOIs agree with each other.
It has also been shown that 
the gaps in the light curves that occur when the signals of inner transiting
planets are removed 
have only a minor effect on the detectability of long-period transiting planets \citep{2017AJ....153..180S}. Thus we believe we have counted all the relevant transiting Jupiter-sized planets around the stars in our SE sample. Nevertheless, we discuss possible effects of detection incompleteness in Section \ref{ssec:incomplete}.

\begin{deluxetable*}{rrrccccr}[!ht]
\tablecaption{Long-period Transiting Planets with Inner Transiting Planets.\label{tab:cjdata}}
\tablehead{
\colhead{KIC} & \colhead{KOI} & \colhead{\kepler} & \colhead{radius ($\re$)} & \colhead{orbital period (days)} & \colhead{eccentricity} & 
\colhead{inner KOIs} 
}
\startdata
\multicolumn{2}{l}{(\textit{confirmed KOIs})}\\
{\bf 11709124} &	435	& 154 & $10.26^{+0.34}_{-0.36}$ & $1250^{+490}_{-240}$ & $0.15^{+0.22}_{-0.11}$ & 435.04 ($1.6\,\re$, $3.93\,\days$)\\
&&&&&& 435.06 ($1.4\,\re$, $9.92\,\days$)\\
&&&&&& 435.01 ($4.4\,\re$, $20.5\,\days$)\\
&&&&&& 435.03 ($2.6\,\re$, $33.0\,\days$)\\
&&&&&& 435.05 ($3.3\,\re$, $62.3\,\days$)\\
{\bf 3239945} &		490	& 167 & $9.99^{+0.14}_{-0.15}$ & $1071.2321^{+0.0010}_{-0.0009}$ & $0.06^{+0.13}_{-0.04}$ & 490.01 ($1.5\,\re$, $4.39\,\days$)\\
&&&&&& 490.03 ($1.5\,\re$, $7.41\,\days$)\\
&&&&&& 490.04 ($1.2\,\re$, $21.8\,\days$)\\
{\it 9663113} & 179 & 458 & $9.62\pm0.39$ & $572.382\pm0.006$ & $0.15^{+0.16}_{-0.10}$ & 179.01 ($7.2\,\re$, $20.7\,\days$)\\
{\it 6191521} & 847 & 700 & $9.55^{+0.51}_{-0.48}$ & $1106.238\pm0.006$ & $0.10^{+0.15}_{-0.07}$  & 847.01 ($8.2\,\re$, $80.9\,\days$)\\
{\bf 3218908}	& 1108 & 770 & $8.03^{+0.34}_{-0.31}$	& $1290^{+870}_{-340}$ & $0.15^{+0.19}_{-0.10}$ & 1108.01 ($3.2\,\re$, $18.9\,\days$)\\
&&&&&& 1108.02 ($1.4\,\re$, $1.48\,\days$)\\
&&&&&& 1108.03 ($1.8\,\re$, $4.15\,\days$)\\
\tablebreak
10187159 & 1870 & 989 & $6.39^{+0.19}_{-0.17}$ & $1310^{+640}_{-240}$ & $0.40^{+0.11}_{-0.10}$ & 1870.01 ($2.1\,\re$, $8.0\,\days$)\\
8738735 & 693 & 214 & $5.83^{+0.37}_{-0.27}$ & $1390^{+1190}_{-370}$ & $0.17^{+0.18}_{-0.11}$ & 693.02 ($3.2\,\re$, $15.7\,\days$) \\
&&&&&& 693.01 ($3.0\,\re$, $28.8\,\days$) \\
8636333 & 3349 & 1475 & $5.74^{+0.43}_{-0.39}$ & $2030^{+1490}_{-560}$ & $0.29^{+0.23}_{-0.20}$ & 3349.01 ($2.8\,\re$, $82.2\,\days$)\\
7040629 & 671 & 208 & $3.30\pm0.16$ & $5690^{+4000}_{-2970}$ & $0.22^{+0.22}_{-0.15}$ & 671.01 ($1.5\,\re$, $4.23\,\days$)\\
&&&&&& 671.02 ($1.4\,\re$, $7.47\,\days$) \\
&&&&&& 671.04 ($1.2\,\re$, $11.1\,\days$) \\
&&&&&& 671.03 ($1.4\,\re$, $16.3\,\days$) \\
{\it 5351250} & 408 & 150 &  $3.3^{+0.19}_{-0.17}$ & $637.21\pm0.02$ & $0.14^{+0.23}_{-0.10}$ & 408.04 ($1.2\,\re$, $3.43\,\days$)\\
&&&&&& 408.01 ($3.3\,\re$, $7.38\,\days$) \\
&&&&&& 408.02 ($2.7\,\re$, $12.6\,\days$) \\
&&&&&& 408.03 ($3.1\,\re$, $30.8\,\days$) \\
&&&&&& 408.05 ($2.0\,\re$, $93.8\,\days$, FP)
\tablebreak
\multicolumn{2}{l}{(\textit{candidate KOIs})}\\
{\bf 5942949}  & 2525	 & \nodata & $11.64^{+2.14}_{-0.65}$ & $1550^{+1010}_{-280}$ & $0.26^{+0.21}_{-0.13}$ & 2525.01 ($1.9\,\re$, $57.3\,\days$)\\
{\it 3558849} &	4307 & \nodata & $10.19^{+0.40}_{-0.35}$ & $1650^{+710}_{-260}$ & $0.54^{+0.09}_{-0.08}$ & 4307.01 ($3.4\,\re$, $161\,\days$)\\
\rule{0pt}{3ex}
10525077 & 5800  & 459\tablenotemark{$^\dagger$} & $5.80^{+0.29}_{-0.25}$ & $427.040^{+0.005}_{-0.004}$ & $0.46^{+0.14}_{-0.09}$ & 5800.01 ($1.5\,\re$, $11.0\,\days$)
\enddata
\tablecomments{The radius, orbital period, and eccentricity of the long-period planets from the analysis of transit light curves of \citet{2019AJ....157..218K}. The radii of the inner KOIs have been updated from the catalog values following the procedure in Section \ref{sec:sesample}.
}
\tablenotetext{\dagger}{The star has a \kepler\ number because the long-period transiting planet listed here has been validated by \citet{2015ApJ...815..127W}; its period could be twice as long as the value here because of the data gap in the middle of the two detected transits. The inner candidate KOI-5800.01 has not been confirmed, and so we included this star in the candidate SE sample.}
\end{deluxetable*}


\section{Results}\label{sec:results}

\begin{figure*}[h]
\epsscale{1.16}
\plottwo{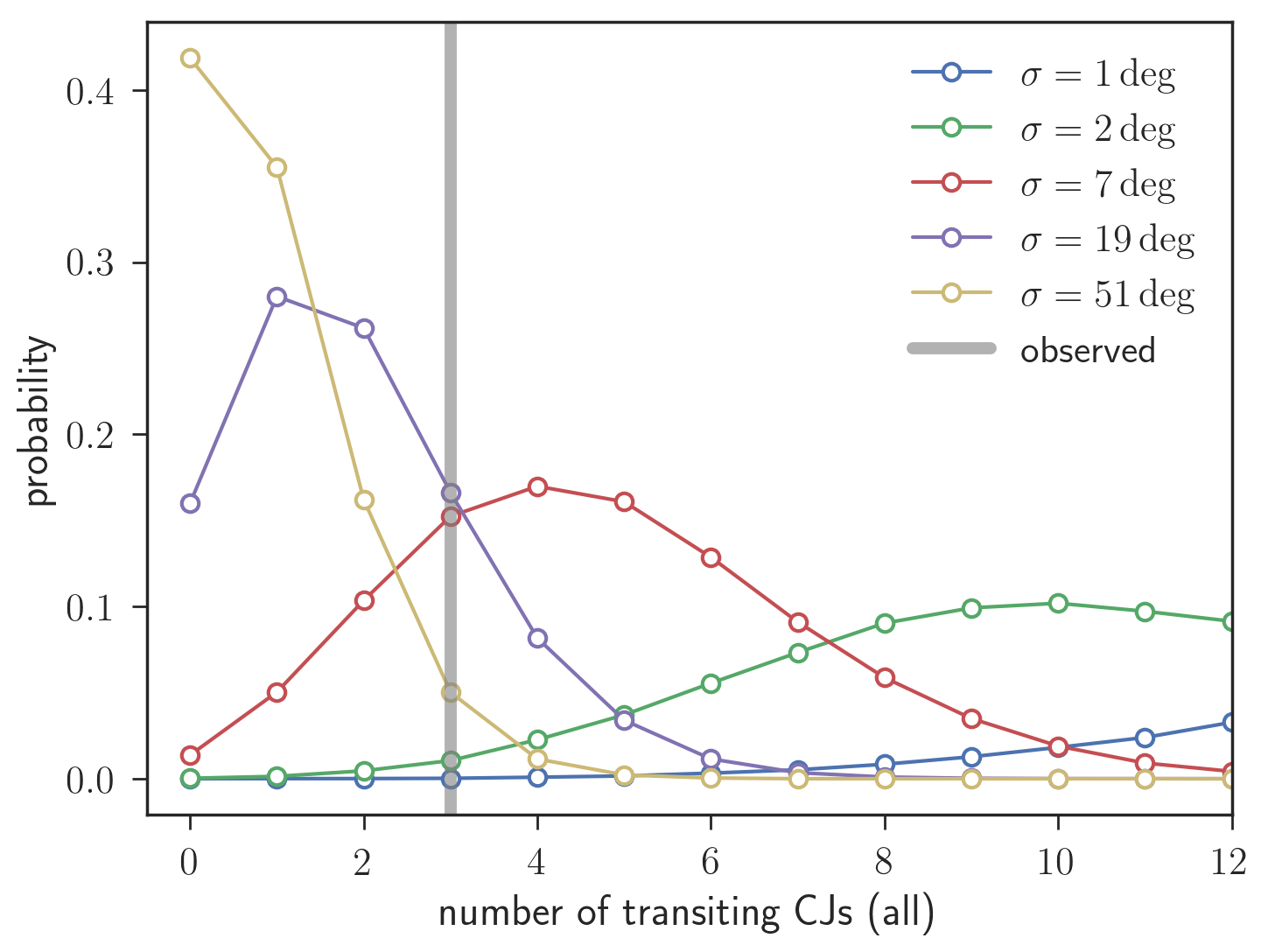}{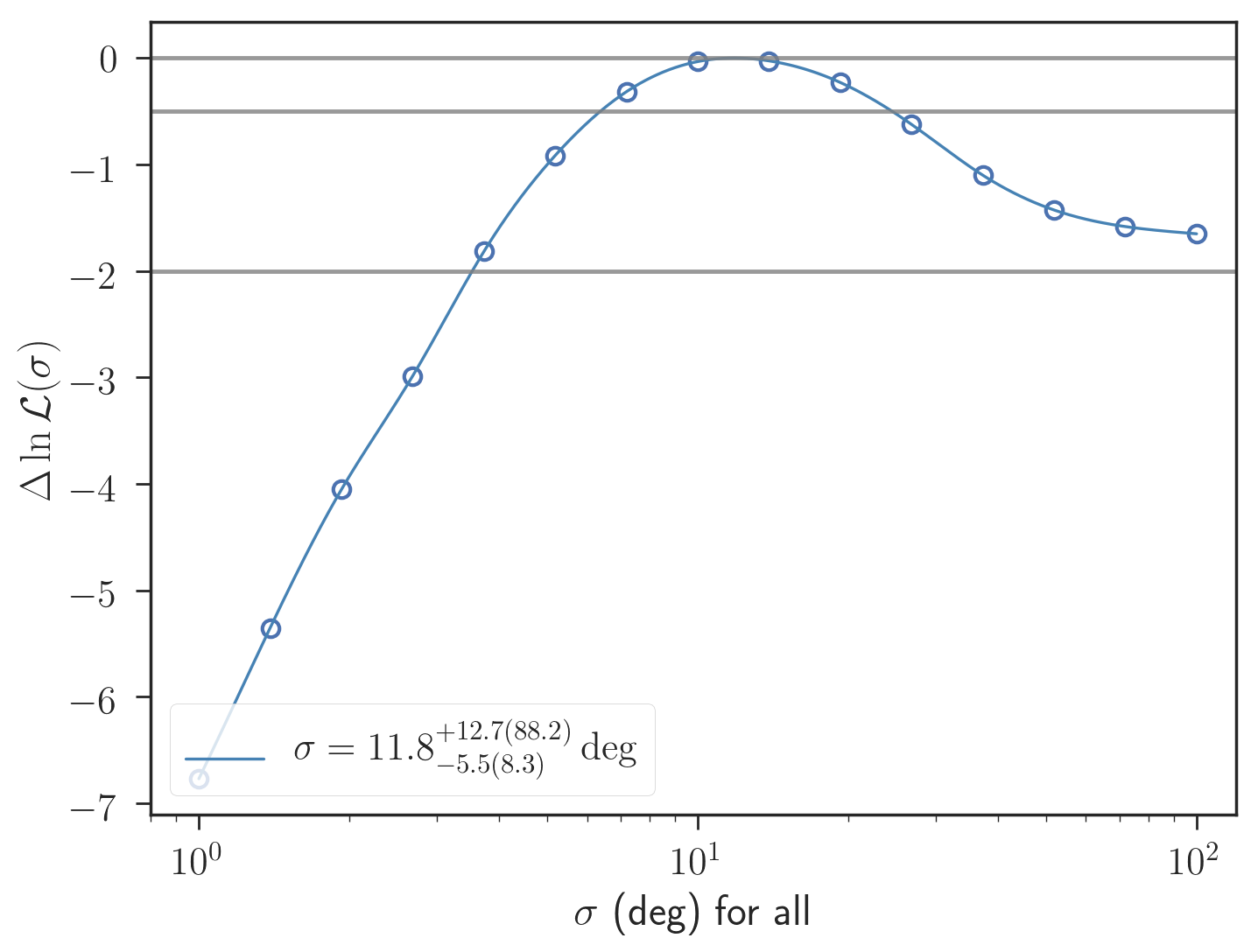}
\plottwo{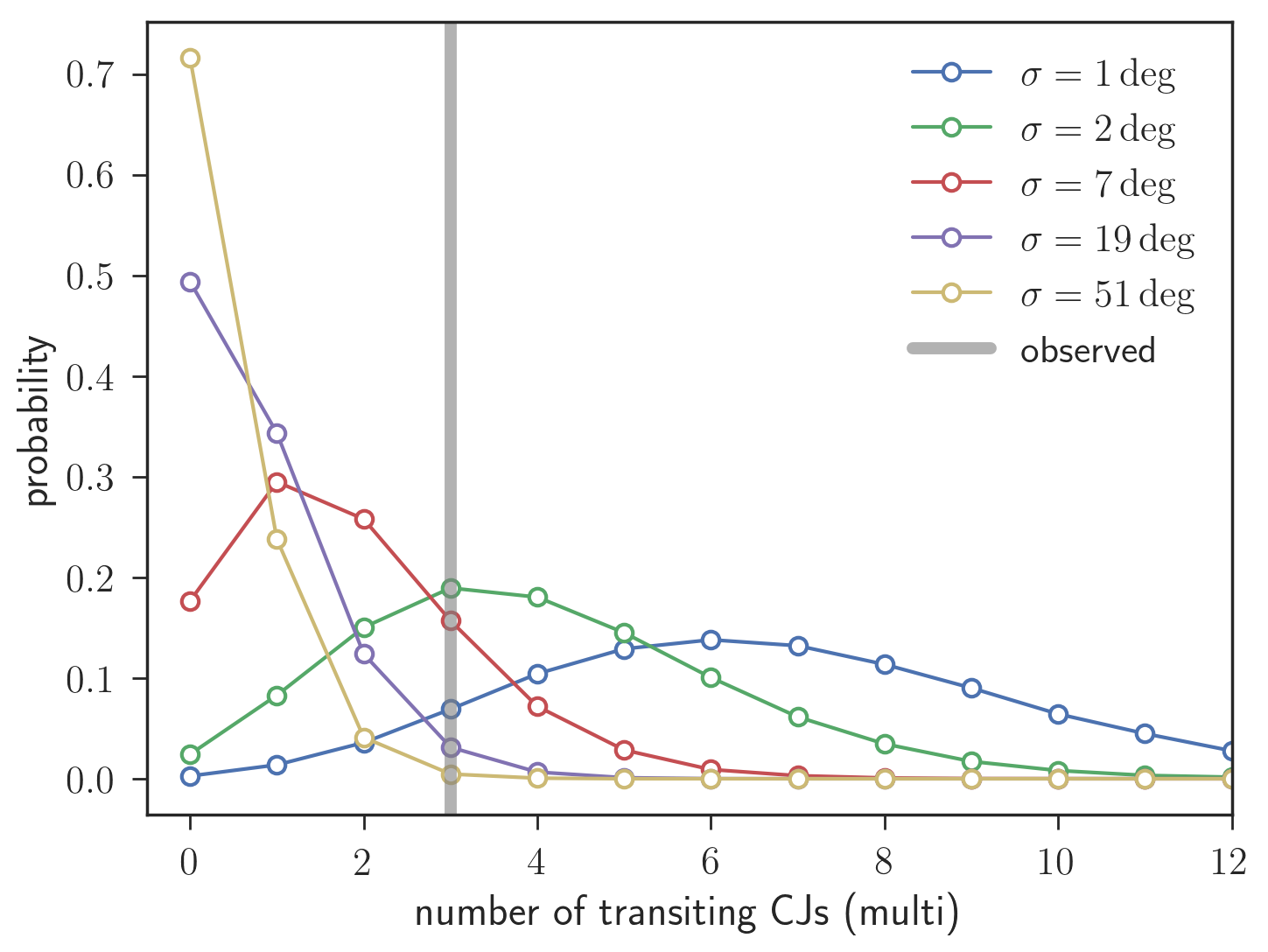}{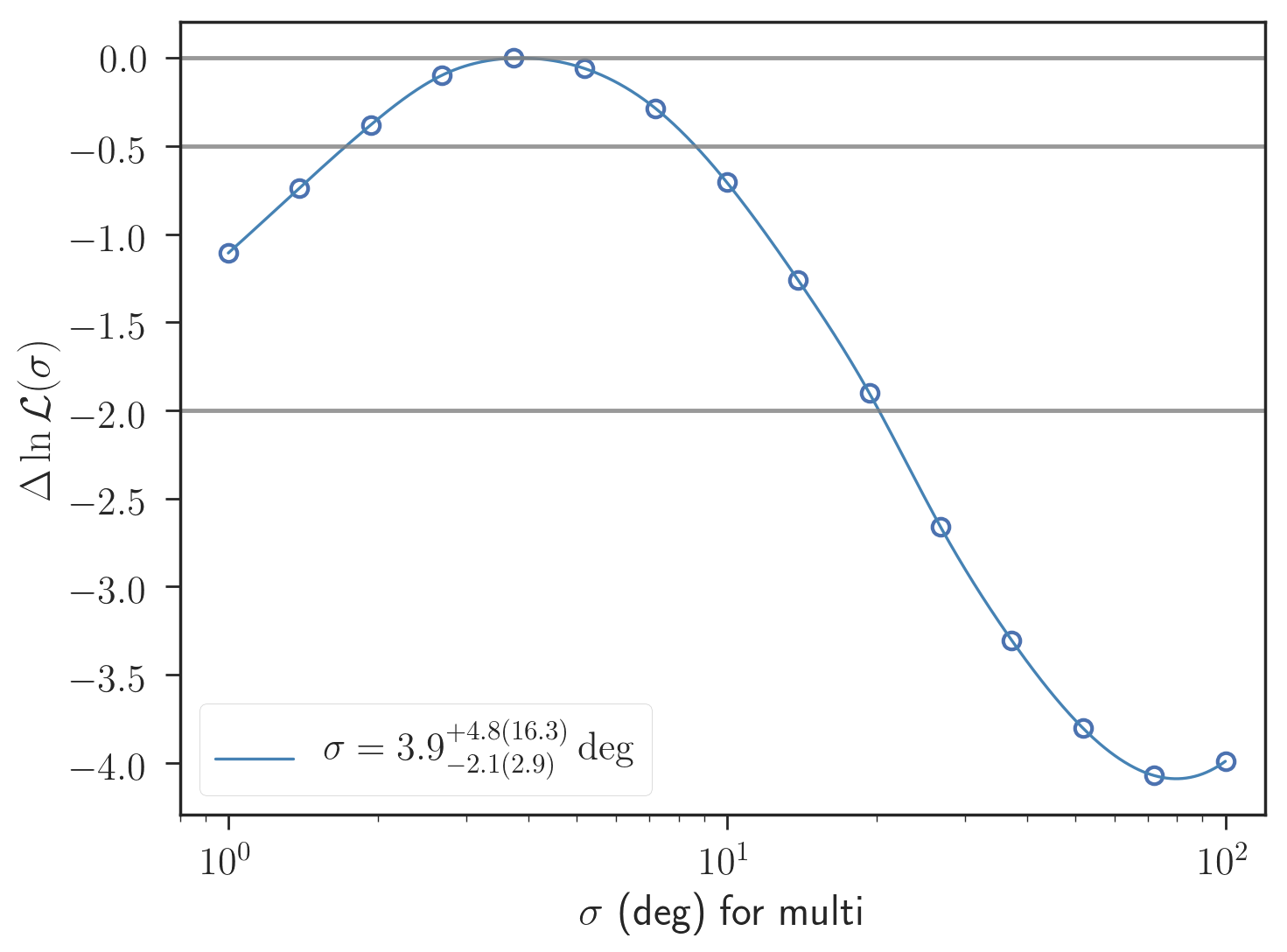}
\plottwo{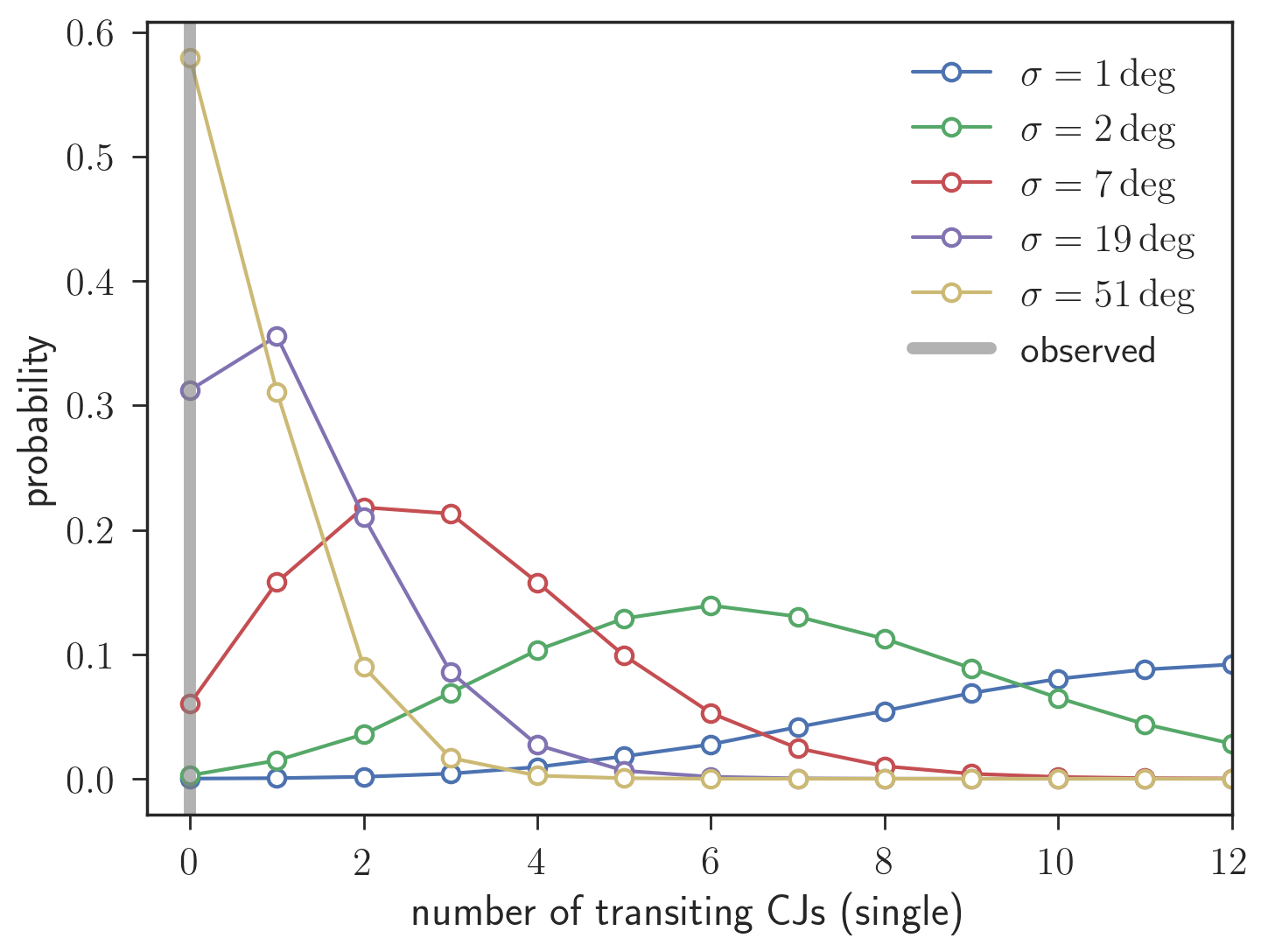}{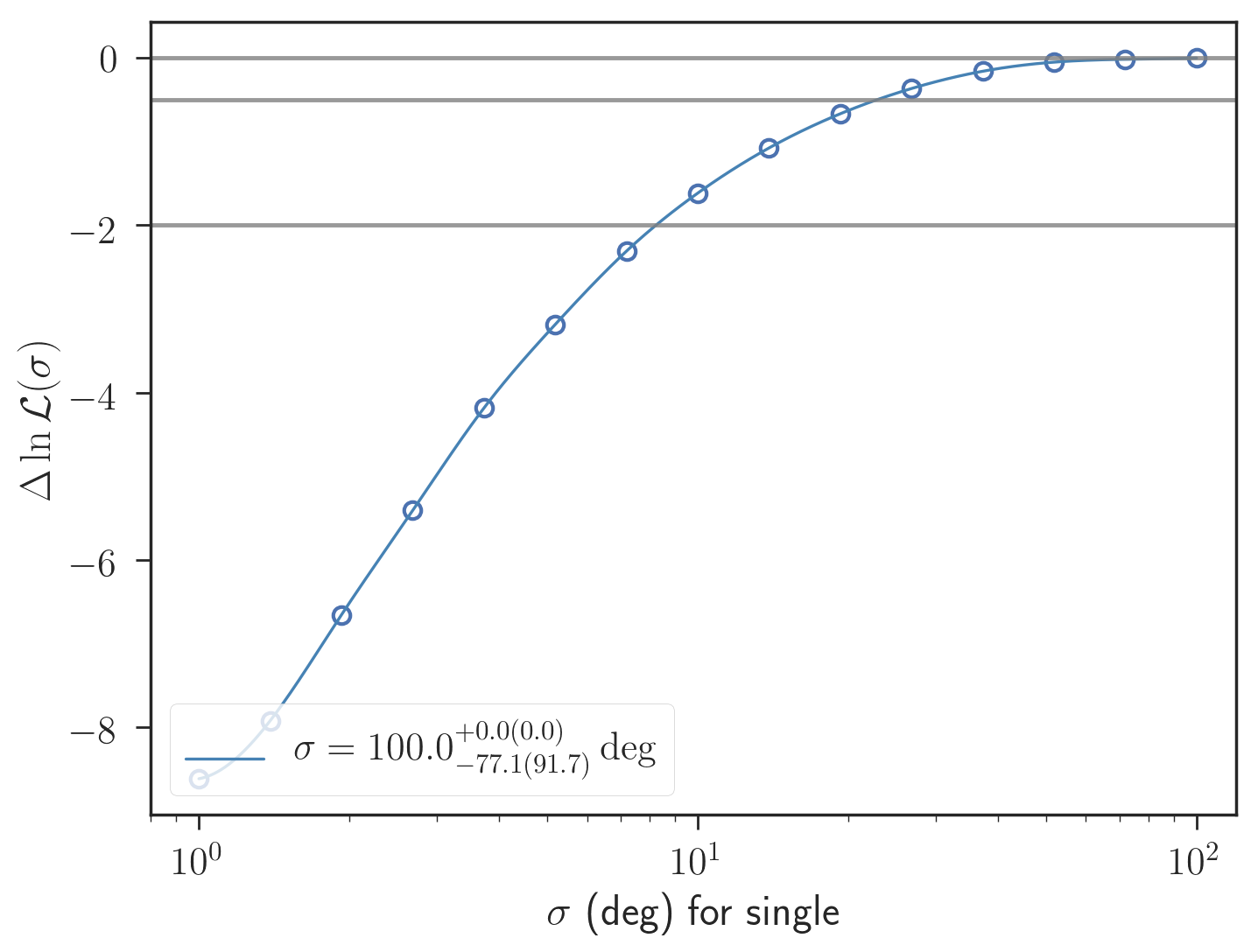}
\caption{{\it Left} --- $p(n_{\rm tCJ}|\sigma)$ for the whole confirmed SE sample (top), subset with transit multis (middle), and that with transit singles (bottom). {\it Right} --- The log-likelihood $\ln \mathcal{L}(\sigma)$ relative to the maximum value. Circles are values from simulations, and the solid lines interpolate them.}
\label{fig:results}
\end{figure*}

Using the CJ occurrence from Section \ref{sec:occ_doppler} and the
transit sample defined in Section \ref{sec:sesample}, we computed
$\mathcal{L}(\sigma)$ for 15 values of $\sigma$ spanning
the range from $1\,\mathrm{deg}$ to $100\,\mathrm{deg}$, spaced equally in the
logarithm of $\sigma$.  For $\sigma\lesssim10\,\mathrm{deg}$, the model
inclination distribution is essentially a Rayleigh distribution, while
for $\sigma=100\,\mathrm{deg}$, the distribution is nearly isotropic.
We interpolated the values of $\mathcal{L}(\sigma)$ using a spline
function, then determined the maximum-likelihood value of $\sigma$ as
well as the intervals corresponding to
$\Delta\ln\mathcal{L}(\sigma)=-1/2$ and $-2$.  Table~\ref{tab:results}
gives the results.  The top panels of Figure \ref{fig:results} show
the functions $p(n_{\rm tCJ}|\sigma)$ for various choices of $\sigma$ (left), along with the log-likelihood function
$\ln\mathcal{L}(\sigma)\equiv p(n_{\rm tCJ}=n_{\rm tCJ, obs}|\sigma)$ (right).

The maximum-likelihood value of $\sigma$ is close to $10\,\mathrm{deg}$,
consistent with the approximate derivation given in Section
\ref{sec:simple}.  The 95\%-confidence lower limit is $3.5^\circ$,
which is somewhat larger than the typical mutual inclinations that
have been inferred for compact multi-transiting systems
\citep[$\sigma\sim$ a few degrees;][]{2014ApJ...790..146F,
  2012ApJ...761...92F}.  Had $\sigma$ been equal to $2^\circ$, for
example, we would have most likely detected 10 transiting CJs (see the
green curve in the top left panel of Figure \ref{fig:results}).

\begin{deluxetable}{l@{\hspace{1cm}}cc}[!ht]
  \tablecaption{Values of $\sigma$ (degrees) for the confirmed SE sample.\label{tab:results}}
\tablehead{
\colhead{} & \twocolhead{confirmed SE\qquad} 
\\
\multicolumn{1}{l}{} & \colhead{Max.\ Likelihood} & \colhead{95\% conf.\ limit}
}
\startdata
all stars ($n_{\rm in}\geq1$)	& $11.8^{+12.7}_{-5.5}$ & $>3.5$ \\
transit multis ($n_{\rm in}>1$)	 & $3.9^{+4.8}_{-2.1}$ & $<20$ \\
transit singles  ($n_{\rm in}=1$) & $>23$ & $>8.3$ \\
$n_{\rm in}\geq3$ & $<2.5$ & $<6.6$ \\
$n_{\rm in}=1$ or $2$ & $>27$ & $>11$ \\
\enddata
\tablecomments{$n_{\rm in}$ is the number of all transiting planets (confirmed and candidate KOIs) with
periods shorter than $130\,\days$.
Lower limits are reported when the interval includes $\sigma= 100\,\mathrm{deg}$.
Upper limits are reported when the interval includes $1\,\mathrm{deg}$.}
\end{deluxetable}

The middle and lower panels of Figure \ref{fig:results} show the
results for the subset of $\mathcal{S}$ with 
multiple transiting inner planets including SEs (``multis''),
and the subset with 
one transiting SE (``singles'').
Inspection of these results shows that the relatively
large $\sigma$ inferred for the whole sample is mainly driven by the
subset of singles, while the multis favor lower values of $\sigma$.
When only the singles are analyzed, the null detection of any
transiting CJs implies $\sigma>8\,\mathrm{deg}$ with 95\% confidence.
This reflects the fact that the upper limit on the CJ transit
probability inferred from the null detection is comparable to the
value expected for random orbital orientations (Section
\ref{sec:simple}).  This value of $\sigma$ is larger than the
mutual inclinations that have have been inferred for
shorter-period multi-transiting systems, as well as those of the Solar System planets. In fact, the same is also
true for the subset of $\mathcal{S}$ having only one {\it or two}
transiting planets, because no transiting CJs were found in that
subset, either.  The result in this case is $\sigma>11\,\mathrm{deg}$.

On the other hand, when analyzing only the multis, the most likely
value of $\sigma$ is $3.9^\circ$, again consistent with the derivation in Section \ref{sec:simple}. This result follows from the
detection of three transiting CJs around stars with multiple
transiting SEs, which is an order of magnitude more than one would
expect if the mutual inclinations were random.  Since all three of the
detections were around stars with three or more transiting SEs, the
results suggest that the mutual inclinations between CJs and
high-transit-multiplicity (i.e., 3 or more) systems of SEs are less than a few degrees, comparable
to the mutual inclinations in the SE system and the Solar System.


\section{Tests for Robustness}
\label{sec:tests}

\subsection{What About Candidate Transiting SEs?}\label{ssec:candidates}

How do the results change if we include candidate transiting SEs,
rather than accepting only confirmed transiting SEs?  If we admit
candidate SEs to the sample, the total number of stars in $\mathcal{S}$
grows to 1{,}721, and the total number of transiting SEs becomes
2{,}335.  Of the stars, 1{,}257 have only one detected transiting SE,
and 464 stars have multiple transiting planets including SEs.  

This enlarged sample includes one clear case of a transiting cold
Jupiter around KIC\,5942949 \citep[][Table \ref{tab:cjdata}]{2016ApJ...822....2U}.  There is also one borderline case: KIC\,3558849.  The star has a transiting cold
Jupiter, but the inner planet candidate in this system (KOI-4307.01) does not quite
fit the definition of an inner super-Earth: its period of 161~days is
longer than the cutoff value of 130~days.\footnote{The SE candidate was reported by \citet{2016ApJ...822...86M} to have a $67\%$ false-positive probability because the transit light curve is V-shaped instead of having a flat bottom.  However, 
this may partly be due to possible long-term TTVs as reported by \citet{2019AJ....157..171K}. The existence of TTVs not only increases the probability that the planet is genuine, but also helps to explain why the transit profile was thought to be V-shaped: unmodeled TTVs tend to make the phase-folded light curve appear to have longer ingress and egress durations.}
However, the period is not too much larger than the cutoff value, and it is within the range of $<400\,\mathrm{days}$ adopted by \citet{2018ApJ...860..101Z}. Because the period cutoff is rather arbitrary,  we decided to analyze both of the cases $n_{\rm tCJ,obs}=4$ and $5$ with and without KIC\,3558849, and confirmed that the results are not sensitive to this difference.

Table \ref{tab:results_cand} gives the results.  The lower limits on
$\sigma$ for transit singles (and doubles)
are slightly weaker, but are similar to the results obtained from the
confirmed-only sample.  This is because the increase in the number of
detected transiting CJs around transit singles (one or two) is
consistent with the increase in the sample size. The number of stars
with transit multis remained mostly unchanged, and so $\sigma$
increased only slightly.

\begin{deluxetable*}{l@{\hspace{1.cm}}cc@{\hspace{1.cm}}cc}[!ht]
\tablecaption{Values of $\sigma$ (degrees) for the confirmed~$+$~candidate SE sample.\label{tab:results_cand}}
\tablehead{
\colhead{} & \twocolhead{including KIC\,3558849\qquad} &  \twocolhead{not including KIC\,3558849\qquad}
\\
\multicolumn{1}{l}{} & \colhead{Max.\ Likelihood} & \colhead{95\% conf.\ limit} &
                       \colhead{Max.\ Likelihood} & \colhead{95\% conf.\ limit}
}
\startdata
all stars ($n_{\rm in}\geq1$)	& $12.1^{+9.0}_{-4.6}$ & $[4.5,42.6]$ & $15.7^{+13.0}_{-6.6}$ & $>5.4$\\
transit multis ($n_{\rm in}>1$)	& $4.8^{+5.6}_{-2.6}$ & $<24$ & $4.8^{+5.6}_{-2.6}$ & $<24$\\
transit singles  ($n_{\rm in}=1$) & $>12$ & $>6.2$ & $>19$ & $>9.1$\\
$n_{\rm in}\geq3$ & $<2.6$ & $<6.9$ & $<2.6$ & $<6.9$\\
$n_{\rm in}=1$ or $2$ & $>15$ & $>7.9$ & $>24$ & $>11$\\
\enddata
\tablecomments{$n_{\rm in}$ is the number of all transiting planets (confirmed and candidate KOIs) inside $130\,\days$. Lower limits are reported when the interval includes $\sigma= 100\,\mathrm{deg}$. Upper limits are reported when the interval includes $1\,\mathrm{deg}$.}
\end{deluxetable*}

\subsection{What If Some Transiting CJs Were Missed?}\label{ssec:incomplete}

We believe that our sample of transiting CJs is complete: the results
of both automated searches and visual inspection were in agreement for
this population, and their signals are 10 or more times larger than
those from the inner planets that were also detected around these
stars. Nevertheless, we decided to check on how serious the systematic
error might be if the sample were incomplete.

We repeated our analysis of the confirmed sample, after artificially
reducing the detectability of transiting CJs to be $60\%$ in step 3 of
Section \ref{sec:framework}. This choice was based on the completeness estimate for
Jupiter-sized planets using the automated pipeline
\citep{2016AJ....152..206F, 2019AJ....157..248H}, and is conservative considering that the orbital periods of actual transiting CJs will be at the shorter end of the range they considered ($\approx$2--20\,yr). 
 Table \ref{tab:results_60} shows the results: the values of $\sigma$ are
generally smaller, because a larger number of transiting CJs is
allowed.  Nevertheless, the basic conclusions are the same.  This is
because our constraint on $\sigma$ relies on the observed
5--15$\times$ boost in the transit probability of cold Jupiters above
the probability expected for uncorrelated orbits.  This makes the
results insensitive to $\mathcal{O}(10\%)$ changes in the number of
detections.

\begin{deluxetable}{l@{\hspace{1cm}}cc}[!ht]
\tablecaption{Values of $\sigma$ (degrees) Inferred for the Confirmed SE Sample, assuming that the detection completeness for transiting CJs is $60\%$.\label{tab:results_60}}
\tablehead{
\colhead{} & \twocolhead{confirmed SE\qquad} 
\\
\multicolumn{1}{l}{} & \colhead{Max. Likelihood} & \colhead{$95\%$ conf. limit}
}
\startdata
all stars ($n_{\rm in}\geq1$)	& $6.9^{+7.4}_{-3.4}$ & $[1.4, 33.4]$ \\
transit multis ($n_{\rm in}>1$)	 & $<4.9$ & $<11.8$ \\
transit singles  ($n_{\rm in}=1$) & $>17$ & $>5$ \\
$n_{\rm in}\geq3$ & $<2.0$ & $<4.9$ \\
$n_{\rm in}=1$ or $2$ & $>21$ & $>7$ \\
\enddata
\tablecomments{$n_{\rm in}$ is the number of all transiting planets (confirmed and candidate KOIs) inside $130\,\days$. Lower limits are reported when the interval includes $\sigma= 100\,\mathrm{deg}$. Upper limits are reported when the interval includes $1\,\mathrm{deg}$.}
\end{deluxetable}

\subsection{Are CJs More Common Around Transit Multis?}\label{ssec:cjocc_multiplicity}

We have assumed that the intrinsic occurrence rate of CJs is the same
for all stars with transiting SEs, regardless of whether there is only
a single transiting SE or multiple transiting SEs.  We found the
occurrence of transiting CJs around stars with a single transiting SE
to be lower by a factor of $\gtrsim5$ than the occurrence around stars
with multiple transiting SEs.  The results indicate that the mutual
inclination between the CJ and the SEs is lower when there are multiple transiting SEs.  However, if the intrinsic occurrence rate of CJs
around stars with only a single transiting SE is $5\times$
lower than the rate of CJs around stars with multiple transiting SEs,
then we would see fewer transiting CJs around transit singles even if
the mutual inclination distribution were the same for all systems.

The SE samples of \citet{2018AJ....156...92Z} and
\citet{2019AJ....157...52B} include stars with both single- and
multi-transiting SEs.  Although the subsamples are small, a difference
in occurrence by a factor of 5 is disfavored by the data.  In the
transit sample of \citet{2018AJ....156...92Z}, $7/22=32\%$ of the SE
systems have CJs. The fraction for inner transit singles is
$5/10=(50\pm14)\%$, and that for inner transit multis is
$2/12=20^{+12}_{-9}\%$. At face value, then, the transit singles could
have a {\it higher} occurrence of CJs, and our results would require
an even larger difference in the mutual inclinations between transit
singles and multis --- although the difference is less than the two-sigma level.
In the Doppler sample of
\citet{2018AJ....156...92Z}, for which the intrinsic multiplicity
(rather than transit multiplicity) is known, the CJ fraction around
single SEs is $9/28=(33\pm7)\%$, and that around multis is
$3/11=(30^{+13}_{-12})\%$. There is no evidence for any difference.

We also note that transit singles and multis generally have similar
properties in terms of orbital periods, planetary radii, and stellar
properties \citep{2018AJ....156..254W}. In addition, similar fractions
of them show TTVs, suggesting that there is not a very large
difference in their intrinsic multiplicities including non-transiting
planets \citep{2018ApJ...860..101Z}. These observations support the
assumption that the transit singles and multis share the same basic architecture except for mutual
inclinations. The preceding comparison of the conditional CJ occurrences further
supports this notion.

Finally, we checked for any differences in the metallicity
distribution between the stars with a single transiting SE
and those with multiple transiting SEs. This is because
high metallicity is strongly associated with the occurrence
rate of CJs. As shown in Figure \ref{fig:comparison_hist},
the stars in our sample with multiple transiting SEs do have a slightly higher mean
metallicity than those with a single transiting SE, but only by $0.03$--$0.04$~dex.
If the occurrence of CJs scales as $10^{2\mathrm{[Fe/H]}}$
\citep{2005ApJ...622.1102F},
the difference in metallicity would
lead to a difference of $\lesssim20\%$ in the occurrence rates of CJs.
This is too small to explain the observed contrast of $\gtrsim500\%$.

\subsection{Do Some of Single-transiting SE Systems Have Undetected Transiting SEs?}

The stars we have classified as transit singles could really be
multi-transiting systems for which some of the transiting planets were
not detected due to a low signal-to-noise ratio.  This would cloud any
distinction between transit singles and multis.  Specifically, this
effect would increase the apparent fraction of stars with transit
singles relative to transit multis and could result in a lower
apparent occurrence of transiting CJs around transit singles. If half
of the observed transit singles are actually transit multis, for
example, the numbers of stars with transit singles and multis in our
sample reverse, and the inferred occurrence rates of transiting CJs
among them would become less different. This is an issue for any comparison
between transit singles and multis. Note, though, that here we are
referring only to {\it transiting} planets that were missed, not about
the planets that exist but do not transit.

To fully resolve this problem, we would need to observe the stars with
a lower noise level, which is impractical.
Nevertheless, we can place an
empirical upper limit on the size of the problem. We wish to
evaluate the probability $\mathcal{P}$ that an observed transit single is
actually a part of a transit multi:
\begin{align}
	\notag
	\mathcal{P} &\equiv p({\rm true\ transit\ multi}\,|\,{\rm obs.\ transit\ single})\\
	\notag
	&={p({\rm true\ transit\ multi})\over p({\rm obs.\ transit\ single})}\\ 
	\notag
	& \quad \times p({\rm obs.\ transit\ single}\,|\,{\rm true\ transit\ multi})\\
	\notag
	&={p({\rm obs.\ transit\ multi})\over p({\rm obs.\ transit\ single})}\\ 
	& \quad \times {p({\rm obs.\ transit\ single}\,|\,{\rm true\ transit\ multi}) \over p({\rm obs.\ transit\ multi}\,|\,{\rm true\ transit\ multi})}.
	\label{eq:pm_s}
\end{align}
Here ``true" refers to the actual (but unknown) transit
multiplicity, including the transiting planets that have not been
detected; and ``obs." refers to the observed transit multiplicity, i.e.,
based on the number of detected transiting planets.
In the last equality, we used the fact that the observed transit multis are subsets of true transit multis, i.e.,
$p({\rm obs.\ transit\ multi}\,|\,{\rm true\ transit\ multi})$ equals $p({\rm obs.\ transit\ multi})/p({\rm true\ transit\ multi})$.

We need to evaluate the last term in Equation~\ref{eq:pm_s}, which is the branching ratio for a
true transit multi to be misclassified as a transit single: we will
denote this ratio as $\mathcal{R}({\rm true\ transit\ multi})$.
Although we do not know the
architecture of true transit multis, we can place an empirical limit on
$\mathcal{R}$ based on the {\it observed} sample of transit multis.
The {\it observed} transit multis are 
no less likely to be observed as transit singles than true transit
multis, because the observed transit multiplicity is always no more
than the true transit multiplicity. 
Symbolically, $\mathcal{R}({\rm true\ transit\ multi}) \leq \mathcal{R}({\rm obs.\ transit\ multi})$.\footnote{Strictly speaking, systems with lower transit multiplicity could lead to lower $\mathcal{R}$ than those with higher transit multiplicity if we consider the case where {\it all} the transiting planets are missed in a system. For example, if either zero or two transiting planets are always missed, for true transit doubles $\mathcal{R}=0$, while for true transit triples $\mathcal{R}>0$.
However, our simulations using observed transit multis verify that such cases are rare: we find that $\mathcal{R}$ does decrease with increasing transit multiplicity, as naively expected.}
In this way, we can obtain an upper limit
on $\mathcal{R}$ using only the observed transit singles and multis.

We evaluate $\mathcal{R}({\rm obs.\ transit\ multi})$ by randomly
assigning planets in the observed transit multis to stars hosting
transit singles, and by counting how many of them would have been
recovered by the \kepler\ transit-finding pipeline based on the transit signal to noise, 
following the procedure of \citet{2017AJ....154..109F}.  Based on 1{,}000 Monte
Carlo realizations, we found $\mathcal{R}({\rm obs.\ transit\ multi})
= (19\pm2)\%$ and $\mathcal{P} \leq (11\pm1)\%$ for the confirmed SE
sample, and $\mathcal{R}({\rm obs.\ transit\ multi}) = (22\pm2)\%$ and
$\mathcal{P} \leq (8\pm1)\%$ for the candidate SE sample.  The more
stringent limit on $\mathcal{P}$ for candidates is due to the larger
fraction of transit singles in the candidate sample ($71\%$) than in
the confirmed sample ($62\%$). We conclude that the sample of transit
singles contains fewer than 10\% of transit multis masquerading
as transit singles. This difference is at most comparable to the
difference in the fraction of transit singles between our confirmed
and candidate SE samples, which yielded consistent results. Thus, the
potential misclassification of transit multis as transit singles is
unlikely to introduce a major bias in our inference. The same is
likely true for other comparisons between transit singles and multis.

\subsection{What About Stars with Multiple CJs?}\label{ssec:multicj}

The occurrence rate for CJs that we adopted for our analysis does not
fully take into account the possible multiplicity of CJs.  Indeed,
among the 12 Doppler SE systems with CJs in the sample of
\citet{2018AJ....156...92Z}, four of them have two CJs with outer
periods ranging from 2000 to 6000~days. \citet{2018AJ....156...92Z} derived the
fraction of SE systems with at least one CJ, while
\citet{2019AJ....157...52B} included only the innermost CJ to derive
the occurrence.  We used their results to calculate the probability to
find one or zero transiting CJs in our sample, because no star in our
sample has more than one transiting CJ.

Our justification for this procedure is that even if multiple CJs
exist, the innermost CJ is the most important one for our analysis,
because it has a higher transit probability.  Based on the CJ
multiplicity observed by \citet{2018AJ....156...92Z}, let us assume
that $1/3$ of our modeled CJ population with typical orbital periods
of $1000\,\mathrm{days}$ (cf.\ Equation \ref{eq:f_f19}) also have outer
CJs with typical periods of $3000\,\mathrm{days}$ (the geometric mean of the
detected outer CJs in the sample). Then, taking into account both the
geometric transit probability and the reduction factor of $4\,\yr/P$
due to the finite duration of the prime {\it Kepler} mission, the
fractional contribution to the total number of transiting CJs due to
the outer planets would be on the order of $(1/3)/5 \sim 10\%$, a level of
systematic error which is itself smaller than the uncertainty in the
overall CJ occurrence.

In this light, our decision to consider only the inner CJs appears to
be justified.  Still, it is interesting to think about the sign of the
possible bias. What happens if two CJs exist in reality? The rigorous
answer depends on the period of the outer CJ and the mutual
inclination between the two CJs, for which we have little or no
information. If the two CJs have a high mutual inclination, the
probability to observe at least one transiting CJ becomes larger than
would be estimated by ignoring the outer CJ.  This is because the
orbital planes of the two CJs sweep out a larger area of the celestial
sphere surrounding the star.  In this case, the true mutual
inclination dispersion between the SE and CJ would be larger than we
have inferred.  If, instead, the orbits of the two CJs are aligned,
then the inner CJ is always more likely to transit, and the
probability to find at least one transiting CJ is unchanged by the
existence of the outer CJ.

\subsection{Dilution due to Unresolved Companions}

The {\it Kepler} transit signals were constructed by summing the light
within a small collection of pixels surrounding the intended target
star.  Whenever this collection of pixels includes an unresolved star
(whether a true companion, or a background object), the transit signal
appears to be smaller because of the constant light from the
neighboring object.  The radii of planets in the same system could all
be underestimated by a common factor if this ``transit dilution'' is
not recognized.

This raises some questions regarding the purity and completeness of
our sample.  One question is whether some of the planets we have
classified as inner SEs are actually giant planets.  This is probably
not a serious problem, because planets larger than $4\,\re$ are
intrinsically much rarer than smaller planets, when considering
orbital periods shorter than 1 year.  Another question is whether we
have missed some SE systems because the transit signals have been
diluted out of detectability.  Most of the time, this type of error
would not result in any bias in $n_{\rm tCJ}(\mathcal{S})$ because the
reason for their exclusion is independent of whether or not a
transiting CJ is detected.  The only problematic cases would be those
for which there is enough dilution for an outer cold Jupiter to appear
to have radii smaller than $7.5\,\re$, but not so much dilution that
the inner SEs would be rendered undetectable.

We think that such cases are rare enough to be negligible for our
study, for two reasons.  First, the required range of dilution factors
(and hence the fraction of companions) that meet the conditions
described above is quite limited.  Let us consider the systems in
Table \ref{tab:cjdata} for which the outer planet appears to be too
small to qualify as a CJ; could these outer planets really be giant
planets for which the transit signal has been diluted?  For KIC
5351250 and 7040629, this is unlikely. The radii would need to be
underestimated by a factor of 3, which would imply that the compact
inner systems consists of 3 or more planets with $r\gtrsim4\,\re$ 
--- and such systems must be rare because 
only Kepler-51 \citep{2013MNRAS.428.1077S,2014ApJ...783...53M} and Kepler-31 \citep{2012ApJ...750..114F, 2014ApJ...784...45R} have such planets in the prime mission sample.\footnote{Another such example has been recently reported around V1298 Tau from the {\it K2} sample \citep{2019ApJ...885L..12D}.}
For the other four
stars (KIC 10187159, 8636333, 8738735, 10525077), the inner planets
have $2$--$3\,\re$ and would still qualify as SEs even if their radii
are actually $1.3$--$2$ times larger than they appear, in which case
outer planets would be classified as CJs.  This scenario requires that
the transit depths be diluted by a factor of $1.7$--$4$, and therefore
that the flux of the unresolved object be $0.7$--$3$ times that of the
planet-hosting star.  Assuming that the companion is physically bound
and the luminosity scales as $\sim M_\star^4$ for main-sequence stars,
this translates into a mass ratio of $0.9$--$1.3$: the two stars need
to be nearly twins.  Considering the binary fraction of Sun-like stars
and mass ratio distribution, only $\approx20\%$ of stars might have
such a companion, and in almost all cases it would have been noticed
based on a double-lined spectrum or a luminosity that appears too
high.  Such coincidences would be even rarer for the case of an
unrelated background star.

A second and more subtle reason is that when the determination of the
stellar radius is based on the parallax, effective temperature, and
apparent magnitudes, the estimated planetary radii are robust to
errors due to dilution, especially for the case of twin binaries.  The
planetary radii are based on the observed transit depth $\delta$, and
the stellar radius is based on the \gaia\ parallax $\varpi$, effective
temperature $T_{\rm eff}$, and observed flux $F_\star \propto
R_\star^2 \varpi^2 T_{\rm eff}^4$. Thus the inferred planetary radius
is $\propto \sqrt{\delta F_\star}/(\varpi T_{\rm eff}^2)$. The
combination $\delta F_\star$ represents the {\it absolute} loss of
flux during transits, i.e., the deficit in the number of photons.  It
does not not depend on how many sources are contributing to the
photometric signal.  Another way to put it is that an unresolved
companion causes the drop in relative flux to appear smaller, but also
makes the host star appear brighter and larger.  These two effects
cancel out as long as the transit depth and stellar flux are measured
in the same bandpass (which is true of \kepler\ and \gaia), and the
measurement of the effective temperature is not significantly biased
by the presence of the unresolved companion (which is true when the
effective temperature is based on spectroscopy, or when the stars are
nearly twins).

\section{Comparison with \citet{2019AJ....157..248H}}\label{sec:compare}

\citet{2019AJ....157..248H} recently presented the results of an
independent effort to determine mutual inclinations between the inner
and outer parts of planetary systems.  While our approach is based on
counting transiting outer planets around stars with transiting inner
planets, they took the opposite approach: counting the cases of
transiting inner planets around stars with transiting outer planets.
They performed the search for long-period transiting planets around
Sun-like stars observed by \kepler\ using the automated pipeline by
\citet{2016AJ....152..206F}, with refined stellar parameters from
\gaia\ DR 2 and their own detrending of the light curves. Their sample
includes 13 stars with long-period giant planets, of which five also
have inner transiting systems consisting of SEs.
\citet{2019AJ....157..248H} pointed out that
this relatively high occurrence of transiting SEs around stars with
transiting CJs is consistent with a picture in which almost all
long-period giant planets are associated with inner compact systems of
smaller planets, with a typical mutual inclination of
$4^\circ$ between the inner and outer systems.
Our analysis led to the inference of a somewhat higher mutual inclination of $11.8^{+12.7}_{-5.5}$~degrees. But it is difficult to make a rigorous comparison between the two analyses, for several reasons.

First, the sample of ``cold giants'' considered by \citet{2019AJ....157..248H} includes planets as small as Neptune, for which the available Doppler data are insufficient to determine
the conditional probability of occurrence around stars with inner SEs. This makes the conditional occurrence rate of SEs interior to their ``cold giants'' more uncertain than that for outer planets that meet our definition of ``cold Jupiters.''

Second, their more inclusive definition of cold giants makes the sample more sensitive to selection effects. If the KOI stars have been searched
more completely for transiting cold giants than the non-KOI stars (as one might expect,
given that the data for KOI stars have lower noise, on average), the occurrence of transiting cold giants with inner planets (i.e., KOIs) would be overestimated, and the mutual orbital inclination would be underestimated. Indeed, if we select the sample using the same criteria but from the catalog of \citet{2019AJ....157..218K}, we find six stars with inner transiting planets among a total sample size of 27 transiting ``cold giants.''  The difference in the number of long-period giant planets arises because of the planets that were not detected by \citet{2019AJ....157..248H}. The resulting fraction $6/27=0.22$ is lower than the fraction $5/13=0.38$ adopted by \citet{2019AJ....157..248H}, and this increases the mutual inclination roughly by a factor of two.

Third, the fraction of transiting cold giants with inner transiting systems 
might be systematically affected by the multiplicity of the inner system, a possibility
which was not taken into account in their analysis.
As we noted in Section \ref{ssec:multicj}, this caveat also applies to our procedure for assigning CJs, but in their case the issue could be more serious because close-in SEs generally have higher multiplicity and smaller orbit spacings than CJs. There is also some evidence that compact multiple-planet systems occasionally have large mutual inclinations \citep{2018ApJ...860..101Z}. In general, any departure from perfectly-aligned orbits acts to increase the probability for {\it at least one} planet to transit, so ignoring this effect could lead to an underestimate of the typical
mutual inclination.

Bearing all these difficulties in mind, we nevertheless tried applying our methodology to the sample of stars with transiting cold giants, adopting the assumption in \citet{2019AJ....157..248H} that all the ``cold giants" have inner SEs. We assigned one inner SE to each star with a period drawn randomly from
the broken power-law distribution inferred for the entire
\kepler\ SEs \citep{2018AJ....155...89P}.  We checked whether the
simulated inner SEs transit or not, and if they did transit, we assumed the
transit signals would always be detected.  We adopted the same model
for the mutual inclination distribution as in Equation
\ref{eq:cosicj}, after swapping $I_{\rm SE}$ and $I_{\rm CJ}$ and
setting $I_{\rm CJ}$ equal to $90^\circ$.  For the 13 stars
discussed by \citet{2019AJ....157..248H}, we found that five
detections imply $\sigma=5.0^{+3.4}_{-1.8}$~deg with
68\% confidence, and $5.0^{+9.8}_{-2.9}$~deg 
with 95\% confidence. 
When we performed the
same exercise for the 27 stars identified by
\citet{2019AJ....157..218K}, we found that six detections imply
$\sigma=9.7^{+5.5}_{-2.9}$~deg with
68\% confidence, and $9.7^{+15.7}_{-5.1}$~deg 
with 95\% confidence. 
These inner
systems include both single- and multi-transiting systems. Thus, we conclude that the
number of transiting SEs found inside transiting long-period giants can be compatible with our analysis of the number of transiting cold Jupiters outside transiting SEs
and the assumption that inner SEs exist around all stars with
long-period planets of Neptune's size or larger.

\section{Summary and Discussion}\label{sec:discussion}

\subsection{Overall Results}\label{ssec:discussion_summary}

Among the sample of stars with close-orbiting super-Earths, there are
too many cases of wide-orbiting transiting cold Jupiters for the
orbital planes of the inner and outer systems to plausibly be
uncorrelated.  The enhancement in the number of outer transiting
planets is about a factor of five for the entire sample of inner
super-Earths, and a factor of 10 or more when the inner system has multiple transiting planets including super-Earths.  Correspondingly, for the sample of
stars with only one transiting super Earth, the number of
detected transiting cold Jupiters is low enough to be compatible with
uncorrelated orbital orientations.

We used these facts to derive constraints on the distribution of
mutual inclinations between inner super-Earths and outer cold
Jupiters.  Specifically, we modeled the distribution as a von
Mises--Fisher distribution and used the data to derive constraints on
$\sigma$, the parameter specifying the width of the distribution.  

As a whole, we found $\sigma=11.8^{+12.7}_{-5.5}\,\mathrm{deg}$.  
Recalling that the CJs around stars with inner super-Earths can account
for all the CJs uncovered from Doppler surveys,
this implies that exoplanetary orbits in systems with CJs are in general dynamically hotter than the planetary orbits in the solar system (even though the solar system does have a ``cold Jupiter'').  This 
seems reasonable, given the broad eccentricity
distribution of CJs that has been observed in Doppler surveys.  
Both the eccentricities and inclinations could be explained as the outcomes
of planet--planet scattering in dynamically unstable systems of
multiple giant planets \citep{2008ApJ...686..580C,
  2008ApJ...686..603J}. Scattering simulations predict that the mean
eccentricity and mean inclination should be of the same order of
magnitude \citep[see, e.g.,][]{2017AJ....153..210H}, and the
inferred mean mutual inclination of our sample $(\pi/2)^{1/2}\sigma=0.26\,\mathrm{rad}$ is comparable to the mean eccentricity $0.3$ of CJs.

We also found that a higher transit multiplicity of the inner system is associated with a lower mutual inclination relative to the cold Jupiter. For example, for inner systems having three or more transiting planets, we found $\sigma$ to be a few degrees, while for inner systems with only one transiting planet, we found $\sigma\gtrsim20\,\mathrm{deg}$. 
Thus, the systems with well-aligned inner orbits are flat across the whole system (at the same level as in the Solar System), and were
apparently unaffected by any dynamical disturbances after their formation.
On the other hand, generic CJs have a broad range of eccentricities (and presumably a broad range of inclinations relative to the initial plane), and thus appear to be dynamically hotter than the undisturbed CJs around multi-transiting inner systems.
This implies that the CJs with higher eccentricities (and inclinations) should generally be associated with inner systems with lower transit multiplicities, some of which are presumably dynamically hot systems with larger mutual inclinations as discussed in Section \ref{sec:intro}.

What could explain this association between the inner and outer systems, spanning an order of magnitude in orbital separation?  It may be that the formation of dynamically hot outer systems facilitates the formation of dynamically hot inner systems, or vice versa.  Alternatively, 
dynamical heating of both the inner and outer systems may result from a common
cause, such as a difference in the protoplanetary disk or stellar environment.
Below, we review some scenarios that have been proposed to explain the dynamically hot inner SE systems with low transit multiplicities, and we discuss how the association between the CJ--SE mutual inclination and inner transit multiplicity might also be explained in each scenario.

\subsection{Relation to the Formation Scenarios of Dynamically Hot Systems of Super-Earths}\label{ssec:discussion_theory}

\subsubsection{Dynamical Heating Due to Outer CJs}

The association between inner and outer systems can be explained if the inner system is dynamically heated due to gravitational perturbations from one or more mutually inclined cold Jupiters. It has been theoretically shown that secular perturbations can decrease the transit multiplicity of an inner system of super-Earths --- either by eliminating some planets through ejections and collisions, or exciting mutual inclinations by driving differential nodal precession of the inner planetary orbits 
\citep{2017AJ....153...42L, 2017AJ....153..210H, 2017MNRAS.464.1709G, 2017MNRAS.467.1531H, 2017MNRAS.468.3000M, 2018MNRAS.478..197P}. 
In these scenarios, the first step
is to raise the inclination (and eccentricity) of the outer planet due to interactions with other giant planets or a wider-orbiting companion star. Then, this dynamical excitation is spread to the inner system until the entire system is dynamically hot. The inner system may also be disrupted by direct interactions with outer giants \citep[e.g.,][]{2017AJ....153..210H, 2017MNRAS.468.3000M}. The cited works have demonstrated that such interactions often produce systems with mutual inclinations $>10^\circ$, which are large enough to explain the distributions we have inferred.

The same type of scenario also naturally explains why the orbits of CJs are better aligned with the inner
system when the inner system has multiple transiting planets.
Even without violent interactions that would reduce the multiplicity of the inner system, an outer giant planet with a modest mutual inclination can disturb the coplanarity of the inner system. For example, \citet{2017MNRAS.468..549B} showed that typical \kepler\ systems with four or more transiting planets are easily perturbed out of multi-transiting configuration by CJs inside $\sim10\,\au$, unless the inner planets are strongly coupled with each other or the outer CJs are on well-aligned orbits.  Such an architecture has been suggested for several multi-transiting systems with non-transiting CJs \citep[e.g., Kepler-48, WASP-47;][]{2017AJ....153...42L, 2017MNRAS.468..549B, 2017MNRAS.469..171R}. Our result suggests that this is the case for general multi-transiting systems with outer CJs. 

\citet{2017MNRAS.467.1531H} performed a population-level study of the effects of outer planets on inner multi-planetary systems. Prior simulations of the multi-planet systems based
on the premise of {\it in situ} assembly \citep{2013ApJ...775...53H}
were able to match the observed properties of \kepler\ multis, but fell short in explaining the observed fraction of transit singles. \citet{2017MNRAS.467.1531H} found that the observed transit multiplicity function of the inner systems can be reproduced if about 40\% of the {\it in situ} assembled inner systems were perturbed by dynamically excited outer giants with eccentricities and inclinations similar to the outcome of scattering simulations by \citet{2008ApJ...686..603J}. The other $\approx60\%$ of the systems remained unperturbed. The required CJ fraction is comparable to the observed rate from Doppler surveys, suggesting that excitation by outer planets could well be an important mechanism for sculpting the observed properties of the inner systems.\footnote{Note, however, that in the simulation of \citet{2017MNRAS.467.1531H}, the presence of multiple outer perturbers turned out to be important for efficiently heating the inner systems. The multiplicity of outer CJs is not yet observationally well constrained, and still needs to be investigated to test the validity of this model.}

\subsubsection{Self-Excitation of the Inner System}

Another proposal is that systems with lower transit
multiplicities originate from ``self-excitation'': dynamical heating due to
planet--planet interactions among the SEs, rather
than a wide-orbiting giant planet.
\citet{2016ApJ...822...54D} and \citet{2016ApJ...832...34M}
studied the excitation during the phase of {\it in situ} assembly and found
that a diversity of dynamical states for compact multi-planet systems
can arise from a corresponding diversity of surface density profiles and gas damping
timescales in protoplanetary disks. While these models can also explain the observed transit multiplicity function,\footnote{It should be noted, though, that
  \citet{2013ApJ...775...53H} could not reproduce the observed fraction 
  of transit singles from their similar simulations, and the reason for
  this difference does not appear to be understood.} the
excited inclinations in these models appear to be generally smaller
(less than $10^\circ$) than the mutual inclination we inferred
for CJs around transit singles and doubles. This is essentially because the
eccentricity/inclination excitation for these close-in planets is limited by 
the strong influence of star's gravity compared to the weak mutual gravitational interactions between low-mass planets.
Similarly, it is difficult to excite $\gtrsim$$10^\circ$
mutual inclinations via longer-term, secular self-excitation after the violent assembly through giant impacts is completed \citep{2012ApJ...758...39J, 2016MNRAS.455.2980B}, because these processes do not produce sufficiently large mutual inclinations to explain the observed fraction of transit singles. Therefore, self-excitation within the inner SE system alone does not seem sufficient to explain the inferred mutual inclination distribution between SEs in low-transit-multiplicity systems and CJs. 
We note, though, that the conclusion depends on the unknown disk properties: for example, excitation of eccentricities and inclinations would be easier in more massive disks with shallower density profiles, in which outer planets have a larger angular momentum deficit that can be transferred to inner planets \citep{2016ApJ...832...34M}. 

\subsubsection{Summary of the Comparison}

The preceding discussion suggests that 
the interactions between the inner and outer systems are essential to explain the inferred mutual inclinations. The facts that CJs are generally dynamically hot, and that most observed CJs are associated with inner SEs, also 
support the conclusion that there exist suitable initial configurations for outer planets to dynamically heat the inner system.

On the other hand, this does not necessarily exclude any contribution from self-excitation, or other possible mechanisms that can excite large mutual inclinations within the inner system. 
Indeed, the current data alone can also be compatible with a scenario in which the observed mutual CJ--SE inclinations simply scale with the mutual inclinations among the inner SE systems.
The model of intrinsic multiplicities and mutual inclinations of the compact \kepler\ systems by \citet{2018ApJ...860..101Z}, as well as other ``two population" models assuming the mixture of coplanar and mutually-inclined systems  \citep{2016ApJ...816...66B, 2018AJ....156...24M, 2019MNRAS.490.4575H}, suggest that the mutual inclination in the ``dynamically hot'' population could be as large as $10^\circ$ or more.
This value itself may be sufficient to explain the large CJ--SE mutual inclinations we inferred for transit singles (and doubles).
Thus, models in which the excitation mechanism does not require CJs are still viable, as long as they can produce sufficiently large mutual inclinations in the inner systems ($\gtrsim$$10^\circ$),
and can explain the association between dynamically hot inner and outer systems. The latter would probably require a large-scale connection of the formation environments in the inner and outer region of the system --- perhaps through a large-scale property of the protoplanetary disk.

Of course, the two scenarios are not mutually exclusive. Self-excitation can happen simultaneously in both inner and outer regions,\footnote{In principle, we do not need to distinguish the two regions. We have adopted this distinction because the region inside $1\,\mathrm{AU}$ is generally devoid of giant planets.} and then the inner systems may further evolve under the gravitational influence of outer giants.   
Self-excitation might also facilitate such evolution, by producing inner systems with wider orbital spacings that are more easily disturbed by outer planets because of their weaker gravitational coupling.

The important difference between the two scenarios, though, is the expected dependence of CJ occurrence on the properties of the inner system. If the CJs are {\it causing} the inner system to be dynamically hot, then dynamically hot systems should always have CJs. This is not required in scenarios such as self-excitation.
This leads us to adopt a working hypothesis that CJs are responsible for {\it all} the dynamically hot inner SE systems and explore its consequences and possible tests.

\subsection{Could the CJs Play the Dominant Role?}


If excitation by outer planets is the main way to form low-transit-multiplicity systems, we would expect the occurrence of CJs to
be higher among stars with a smaller number of transiting SEs.  This
would invalidate an assumption we made in our analysis: we divided
the sample based on inner transit multiplicity and assumed they all had
the same rate of CJ occurrence. However, if CJs are actually more common around
systems with lower transit multiplicities, this would only strengthen our conclusion that
CJs around SE systems with lower transit multiplicities have higher
mutual inclinations --- because we found fewer transiting CJs around such
systems.

This scenario might appear to be contradicted by the lack of any clear
difference in the [Fe/H] distributions of stars with transit singles
and transit multis  \citep{2018AJ....156..254W}, as pointed out by \citet{2018ApJ...860..101Z}.
Because CJs occur more frequently around more metal-rich stars
\citep[e.g.,][]{2005ApJS..159..141V}, if stars with transit singles are preferentially
associated with CJs, they would have relatively
metal-rich hosts.
However, the null result does not
necessarily exclude an association between CJs and
dynamically hot inner systems, considering that the population of
transit singles consists of both dynamically hot systems and flatter
multi-planet systems, with relative abundances that are not well constrained
because of the degeneracy between intrinsic multiplicity and mutual inclination. In fact, \citet{2014ApJ...790...91S} argued that even if 50\% of stars with SEs
also have giant planets, the [Fe/H] distribution of the host stars could
still be statistically indistinguishable from a sample of SE host stars chosen
randomly, without regard to outer giant planets.

We can flesh out this argument with the following construction.  In
the model of \citet{2018ApJ...860..101Z}, the observed transit
multiplicity distribution of \kepler\ systems may be explained if 25\%
of the systems belong to a dynamically hot population. Assume that all
of these dynamically hot systems have CJs.  Then, in order to produce
an overall CJ occurrence of $1/3$ as observed, the CJ occurrence among
the dynamically cold 75\% of the sample needs to be $1/9$, viz.,
$1/3=(1/4)\times1+(3/4)\times(1/9)$.  Next, consider that $2/3$ of
the {\it Kepler} systems are single-transiting systems. This implies
that $(1/4)/(2/3)=3/8$ of the transit singles are dynamically hot
systems, and the other $5/8$ are dynamically cold.  The CJ occurrence
around transit singles is therefore $3/8 +
(5/8)\times(1/9)=4/9=44\%$.
On the other hand, the CJ
occurrence around transit multis, which are all dynamically cold, is
only $1/9=11\%$. These values roughly agree with what has been
observed in the available samples (Section \ref{ssec:cjocc_multiplicity}). Also, the fraction of CJ hosts among transit singles (44\%) is consistent with the current null detection of the difference in [Fe/H] distribution, according to \citet{2014ApJ...790...91S}.

Even if we should not expect any strong metallicity trends, we may
still see some correlations arising from the association between the
CJ occurrence and stellar metallicity. For example, the lack of high
transit-multiplicity systems around metal-rich stars noted by
\citet{2018AJ....156...92Z} could be due to the higher occurrence of
CJs that reduce the transit multiplicity of the inner
system.\footnote{\citet{2018AJ....156...92Z} interpreted this trend as
  the anti-correlation between CJs and the {\it intrinsic}
  multiplicity (rather than the transit multiplicity) of the inner
  system.}  Similar correlations could be seen with any property that
is correlated with the CJ occurrence, such as the stellar mass
\citep{2010PASP..122..905J}. For example, the M-dwarf subsample of
\kepler\ SEs appears to include a smaller fraction of transit singles
\citep{2016ApJ...832...34M} which could be connected to a lower CJ
occurrence rate.

We note that the construction presented above is meant to show that the current data could be compatible with an (extreme) assumption that low-transit-multiplicity systems are entirely produced by external CJs. As we have discussed, the other extreme is that CJs are irrelevant, and the inferred CJ--SE mutual inclination simply mirrors the mutual inclinations of the inner SE system which were excited by other mechanisms. In this case, we would not expect any difference in CJ occurrence around systems with different transit multiplicities. 
This will be one of the key pieces of information to further shed light on the dominant mechanism to produce dynamically hot inner systems. 

\subsection{Future Probes}

To summarize the overall story:
compact systems of SEs exist around about $1/3$ of stars, of which
$1/3$ also have CJs. The mutual inclinations between the orbits of the SEs and the CJs are generally on the order
of $10^\circ$. Those systems that have larger inclinations tend to
have lower transit multiplicities, hinting that the inner systems were dynamically heated by outer planets.
There are several ways that the veracity of this story might be checked
in the future.

Above all, we need more systems where both CJs and inner SEs can be detected and characterized. In particular, it is important to understand the dependence of CJ occurrence on the inner transit multiplicity to understand the role of CJs in shaping the system architecture. Such information may come from ongoing Doppler programs targeting \kepler\ systems \citep[e.g.,][]{2019AJ....157..145M}. Ultimately, \gaia\ data should also provide a large sample of stars which have both an inner system of transiting planets and an astrometrically-detected outer giant planets. 
The sample would also provide better and more direct constraints on the mutual inclination distribution, because the transit and astrometry data will constrain the line-of-sight inclinations of inner SEs and CJs, respectively. 

It would also be of interest to compare the eccentricities of CJs outside transit singles and multis.
If the larger mutual inclinations around transit singles are indeed associated with dynamically hot CJs, rather than self-excited inner systems, we should see larger eccentricities for the CJs around transit singles as well --- although this comparison
may suffer from the same problem that we discussed in the previous
section regarding the comparison of the [Fe/H] distributions.  On the other hand, CJs around high
transit multiplicity systems would generally have low
eccentricities. The transit durations of the three CJs in our sample --- all of which involve
multi-transiting inner systems --- are in fact consistent with low
eccentricities (see Table \ref{tab:cjdata}), although the constraints are not very tight.
Eccentricities may also be useful to test other possible mechanisms,
such as star--planet interactions \citep{2016ApJ...830....5S} and warped disks \citep[e.g.,][]{2018MNRAS.477.5207Z},
that can cause misalignments
between the inner and outer system without violent dynamical excitation.
If such mechanisms are significant, even inclined outer giants could still have circular orbits.

Measuring the obliquities of stars may also be useful as another way
to identify the dynamically hot SE population. Although the
$v\sin i$ distribution of transiting planet hosts does not show clear
evidence that stars with transit singles and multis have different
obliquity distributions \citep{2014ApJ...783....9H,
  2014ApJ...796...47M, 2017AJ....154..270W}, this may simply be due to
the limited sensitivity of $v\sin i$-based comparisons to small differences in
inclination. 
In this regard, it is interesting that a highly oblique star with a single-transiting planet has been revealed by asteroseismology \citep{2019AJ....157..137K}.
More precise measurements with the Rossiter--McLaughlin
effects for a larger number of small transiting planets are warranted.

Although we focused exclusively on inner super-Earths smaller than $4\,\re$ in this paper, some 50\% of warm giants with $10\,\days\lesssim P \lesssim100\,\days$ are known to co-exist with SEs in compact multi-planetary systems, possibly sharing a similar origin to the close-in SEs \citep{2016ApJ...825...98H}. 
Then, our inference might imply that similar misalignments also exist between warm giants and CJs.
Such misalignments have been invoked theoretically to explain the population of eccentric warm Jupiters \citep{2014ApJ...781L...5D} and inferred observationally from the clustering of apsidal orientations of some warm- and cold-giant planet pairs on eccentric orbits \citep{2014Sci...346..212D}.
This possibility may also be tested with the \gaia\ sample, and potentially has broader implications
for the long-term dynamics of giant planets inside CJs. 
For example, misalignments are part of the initial conditions required to explain the large eccentricities of some warm giant planets \citep{2017MNRAS.472.3692A}. 
The excitation of large eccentricities, when combined with tidal dissipation, could also lead to the conversion of a warm giant into a hot Jupiter \citep{2017AJ....154...64M}. A similar mechanism could also be responsible for the formation of hot sub-Neptunes and sub-Saturns, whose occurrence rates are correlated with stellar metallicity \citep{2018PNAS..115..266D, 2018AJ....155...89P}, and hence the occurrence of CJs.\\

\acknowledgments

We thank Dan Foreman-Mackey for helpful conversations in the early stage of this work. We also thank an anonymous referee for careful reading of the manuscript and thoughtful suggestions on the interpretation of the results, which helped us to improve the manuscript. This paper includes data collected by the \kepler\ mission. Funding for the \kepler\ mission is provided by the NASA Science Mission directorate. This work was performed in part under contract with the California Institute of Technology (Caltech)/Jet Propulsion Laboratory (JPL) funded by NASA through the Sagan Fellowship Program executed by the NASA Exoplanet Science Institute. KM gratefully acknowledges the support by the W.\,M.~Keck Foundation Fund.





\bibliographystyle{aasjournal}


\listofchanges

\end{document}